\documentclass[aps,prc,reprint,superscriptaddress,nofootinbib,floatfix]{revtex4-2}

\usepackage{silence}
\usepackage{amsmath,amssymb}
\usepackage{physics}
\usepackage{bm}
\usepackage{dcolumn}
\usepackage{braket}

\usepackage{graphicx}
\usepackage{rotating}
\usepackage{multirow}
\usepackage{array}
\usepackage{longtable}
\usepackage{booktabs}

\usepackage[dvipsnames]{xcolor}

\usepackage[colorlinks=true,allcolors=Blue,pdfusetitle]{hyperref}
\usepackage{orcidlink}

\begin{document}

\title{Quantum Simulation of Nuclear Shell Model Using GCM-Based Methods on NISQ Devices}
\author{Durgesh Pandey\orcidlink{0009-0001-0877-1980}}
\email{durgesh_p@ph.iitr.ac.in}
\affiliation{Department of Physics, Indian Institute of Technology Roorkee, Roorkee 247667, India}

\author{Ashutosh Singh\orcidlink{0009-0001-9202-571X}}
\email{ashutosh5765449@gmail.com}
\affiliation{Department of Physics, Indian Institute of Technology Roorkee, Roorkee 247667, India}
\affiliation{CNRS/IN2P3, IJCLab, 91405 Orsay, France}

\author{Ankit Kumar Das\orcidlink{0009-0001-2252-2657}}
\email{ankitk_das@ph.iitr.ac.in}
\affiliation{Department of Physics, Indian Institute of Technology Roorkee, Roorkee 247667, India}

\author{P. Arumugam\orcidlink{0000-0001-9624-8024}}
\email{arumugam@ph.iitr.ac.in}
\affiliation{Department of Physics, Indian Institute of Technology Roorkee, Roorkee 247667, India}
\affiliation{Centre for Photonics and Quantum Communication Technology, Indian Institute of Technology Roorkee, Roorkee 247667, India}

\date{\today}

\begin{abstract}
Based on the Generator Coordinate Method (GCM), we use a Quantum GCM (QuGCM) within a hybrid quantum-classical framework to simulate low-lying eigenstates of nuclear systems on quantum devices. The generator basis states are constructed from Hartree-Fock (HF) reference states, excited via symmetry-adapted unitary coupled-cluster (UCC) operators. These states are prepared as non-orthogonal quantum circuits and measured pairwise to compute the required overlap and Hamiltonian kernels. The resulting data is processed using a classical generalized-eigenvalue solver, following the GCM formalism, to extract the system's energy spectrum. To enhance efficiency and reduce circuit depth, we apply the Adaptive Generator Coordinate Inspired method (ADAPT-GCIM), which iteratively selects generator excitations based on energy gradients, thereby avoiding the need to explore the full Hilbert space. Our implementation is applied to nuclear systems, specifically the deuteron with the Reid68 potential and shell-model Hamiltonians of $^6$Li and $^{38}$Ar. For each system, both the QuGCM and ADAPT-GCIM methods produce energy spectra in agreement with classical diagonalization results, demonstrating robustness even under noise and limited-depth constraints. Additionally, we compare fermionic encoding strategies, specifically Jordan-Wigner (JW) transformations of one-hot (OH) encoding and Gray code (GC) mappings, and show that GC encoding reduces circuit complexity and improves fidelity during multi-reference state preparation. Our findings indicate that QuGCM and ADAPT-GCIM provide a practical and scalable path toward simulating correlated quantum systems, with lesser vulnerability to noise and better compatibility with the limitations of current quantum hardware.
\end{abstract}

\maketitle

\section{Introduction} \label{Introduction}
The Generator Coordinate Method (GCM) is a variational approach in quantum many-body theory in which the total wavefunction is constructed as a superposition of states labelled by collective coordinates, allowing essential correlations to be incorporated systematically. The core idea behind GCM is to express the wavefunction of a quantum system as a superposition of non-orthogonal reference states, which are usually Slater determinants~\cite{hill1953nuclear}, each depending on a set of continuous or discrete generator coordinates. These coordinates may correspond to nuclear shapes~\cite{Wong1975,Verriere2020}, time dependent fluctuations~\cite{Chattopadhyay1978}, pairing amplitudes~\cite{HilaireZuker1987,Uzawa2024}, cranking frequency~\cite{Ansari1985}, etc. GCM is highly useful and effective beyond mean-field theory by including several constraints over the mean field theory~\cite{Shimada2021}. Therefore, it is well-suited to be used along with self-consistent mean field calculations in Hartree-Fock (HF), Hartree-Fock-Bogoliubov, and Bardeen-Cooper-Schrieffer (BCS) theory. These collective coordinates represent important collective degrees of freedom, such as deformation, vibration, or pairing, that capture critical correlations in the system~\cite{RingSchuckBook, Bender2003, Griffin1957collective}. By mixing these states, the method improves upon the mean-field approximation. The coefficients that control this mixing are determined by solving the Hill–Wheeler equation~\cite{hill1953nuclear, Bender2003, RingSchuckBook}, a generalized eigenvalue problem, which ensures that both ground and excited states can be systematically obtained.
It is one of the important and foundational frameworks for dealing with highly complex quantum systems. It was first introduced by Hill and Wheeler in 1953 to explain Nuclear fission. The method was later simplified to explain the collective model in nuclei by Griffin and Wheeler. Apart from the extension of the method to study clustered nuclei~\cite{horiuchi1970generator}\, Giant Monopole Resonances~\cite{krewald1976selfconsistent} and Super deformations~\cite{dancer1999generator} to beyond mean field models in nuclear systems, the method is also useful for general quantum systems. It has also been extended to multi-electron systems~\cite{trsic2004generator}, like molecular~\cite{jorge2000improved} and atomic systems~\cite{malli1998generator}. This wide application of the method in nuclear systems, along with its generalized theory, is well explained in the literature.  

To meet the rigorous calculations of highly demanding methods or models like GCM, the functional ability of classical computers has always been fruitful for decades. The implementation of GCM using computational techniques involves grid formation that discretizes the generator coordinates space, constructing the corresponding suitable generator states, followed by evaluation of each element in the Hamiltonian and norm matrices among these states, and then finally solving the resulting generalized eigenvalue. This has been implemented for different generator states classically. However, due to limitations in computing resources and efficiency, various techniques, such as the Onishi formula and the Pfaffian formula, have been used to reduce the load of finding each overlap~\cite{Onishi1966, Robledo2009}. The computational compatibility of GCM depends upon numerous factors, such as whether the grid is sufficiently large or not, whether overlap kernels do not become numerically unstable or ill-developed generalized eigenvalue equation. The major factors causing the increase in calculations are an increase in the number of grid points or the number of basis states. By doubling the number of coordinates or grid points, the total basis size increases by $2^d$, where $d$ is the total number of coordinates. Therefore, with an increase in the number of collective coordinates, the mesh size grows exponentially. Even two-dimensional GCM meshes can exceed $10^4$ points, making the Hamiltonian and norm matrix evaluations (scaling as $N^2$) infeasible on typical CPUs of that time. Despite applying constraints and many symmetry rules to reduce the amount of computation needed, the evaluation of computational GCM remains resource- and time-intensive. To overcome the computational burden of these calculations across different generator coordinates, recent developments include the integration of Noisy Intermediate-Scale Quantum (NISQ) devices by developing a hybrid quantum-classical approach.

Quantum Computing is a fundamentally different framework that can encode quantum systems into its own quantum system consisting of qubits and gates, equivalent to a local harmonic oscillator basis along basic operations~\cite{Feynman1982, Lloyd96, NielsenChuang00}. These basic operations can be combined or superimposed over each other to implement any desired operator. This method is highly promising when the expected wavefunction is a superposition over a large basis. This basis must span the possible Hilbert space where the solution lies. Due to limitations in current NISQ devices in terms of the number of qubits and fidelity of quantum operations, mainly hybrid techniques using classical optimization have been in use over the decade~\cite{preskill2018quantum,mcclean2016theory}. The application of a quantum computer is to determine energy spectra of quantum systems such as nuclei, and this is accomplished by constructing a physically realistic and accurate Hamiltonian of the system, and then finding a suitable wavefunction by encoding both in terms of the known qubit basis of the quantum computer. The coefficients of the wavefunction for individual basis states can be determined by two broad approaches.
 
The first approach is the optimization of the wavefunction to get the minimum expectation value with the Hamiltonian. Broadly, coefficients depend on the parameters, which are systematically adjusted to minimize the cost function. These methods, like VQE~\cite{peruzzo2014variational}, etc., have been widely adopted for quantum simulations in physics and chemistry, leading to significant improvements in energy calculations for spectra of atomic, molecular, and nuclear systems~\cite{Complexscalling_ashutosh}. 
Many examples of techniques that use optimization are Variance Quantum Eigensolver (VQE), Adaptive VQE (ADAPT-VQE) \cite{Siwach2021,romero2018strategies,shen2017quantum, Kandala2017, Kandala2019,colless2018computation, Huggins2020, Cao2019,grimsley2019adaptive,grimsley2019trotterized, Verteletskyi2020, McArdle2019, McArdle2020, Tilly2022}, the Quantum Approximate Optimization Algorithm (QAOA) \cite{Farhi2014, Stein2022}, quantum annealing \cite{bharti2021noisy,albash2018adiabatic}, and approaches like Gaussian boson sampling \cite{Aaronson2011}, analog quantum simulation \cite{Trabesinger2012,georgescu2014quantum}, and iterative quantum assisted eigensolvers \cite{Motta2020, Parrish2019, Kyriienko2020}. These methods distribute computational tasks between classical and quantum computers to conserve quantum resources and enable the use of near-term devices. 

Accurately finding the energies of ground and excited states, along with their wave functions, is essential for understanding many physical phenomena in molecules and materials. This includes phenomena such as high-temperature superconductivity in materials like cuprates \cite{Imada1998metal}, chemical reactions that break bonds, and complex electronic processes in catalysts containing transition metals \cite{Witzke2020} or f-block atoms \cite{Gould2022}. The spin, electronic properties, and dynamics of these systems are key to understanding the relationship between structure, properties, and function in fields such as catalysis, sensors, and quantum materials.
However, this task becomes very difficult due to the strong correlation. Traditional wave function methods often fail with these complex cases or become too computationally expensive as the system size grows~\cite{shavitt2009many}. Because of this, while newer variational quantum algorithms have been developed, the highly non-linear way their wave function is set up leads to complex energy landscapes that are hard to optimize. As an alternative, near-term strategies that solve a generalized eigenvalue problem within a constructed subspace are being explored, as they are often better suited for NISQ devices.
  
The other approach is to solve the generalized eigenvalue problem in a chosen basis, which can be non-orthogonal too. This involves direct diagonalization of the Hamiltonian and overlap matrices, often resulting in more robust and NISQ-adapted algorithms, since it typically avoids deep quantum circuits to include all basis at once, as there are no optimization steps. Unlike a direct eigensolver, GCM builds quantum states from physically meaningful collective configurations, giving a compact way to capture correlations in many-body systems. Recent studies have introduced the QuGCM and ADAPT-GCIM techniques, which are based on the GCM. These methods show excellent agreement with energy calculations for different molecular models ~\cite{zheng2023quantum,zheng2024unleashed}. However, as GCM is developed along with the nuclear models, it is highly probable to implement these quantum-classical hybrid techniques in nuclear models. So far, many such integrated hybrid techniques are already in use for nuclear systems ~\cite{Yao2010, Egido2016, Hizawa2021}. These methods are a near-term alternative to VQE and ADAPT-VQE, designed to overcome their limitations. They use low-depth quantum circuits and existing ansatz to explore a quantum subspace, targeting specific energy levels and states \cite{Baek2023}. These algorithms perform efficient quantum subspace diagonalization by directly solving generalized eigenvalue equations, which avoids the extensive parameter optimization required by VQE and similar variational methods. Furthermore, ADAPT-GCIM automates subspace selection using gradient-based strategies. This leads to faster convergence to both ground and excited states, reducing simulation time and improving precision, particularly in strongly correlated systems. 

Our study extends the subspace quantum eigensolver techniques QuGCM and ADAPT-GCIM to the domain of nuclear systems. By employing nonorthogonal quantum subspace expansion and direct eigenproblem solvers, we leverage the robust, NISQ-friendly advantages of these methods, making them well-suited for current and future quantum computing architectures. This approach effectively tackles the optimization challenges of traditional VQE methods by maintaining an adaptive framework. Recent research has focused on practical solutions to mitigate optimization limitations, such as using domain-specific knowledge from classical quantum chemistry to build high-quality wave functions \cite{Hirsbrunner2024} or developing adaptive algorithms that still require optimization. For instance, recent work by Zhang and Lacroix applies ADAPT-VQE to nuclear pairing correlations and proposes a method to find excited states from its convergence path \cite{n-pscatteringDenis,Denisexcitedstatefromadaptvqe}. In contrast, our approach uses conventional Unitary Coupled Cluster (UCC) or subset cluster sets like Unitary Coupled Cluster Singles and Doubles (UCCSD) or UCCD as excitation generators on the basis for subspace expansion, which sets a lower bound on the optimization problem found in VQE.  Building upon the foundational strategies established for molecular systems, we implement an optimization-free, gradient-based automated basis selection method tailored for nuclear environments. This hierarchical adaptive strategy balances subspace expansion and ansatz optimization. These preliminary strategies demonstrate their effectiveness in handling strongly correlated molecular systems by significantly reducing simulation time and enabling deployment on real quantum computers. Building on these insights, we extend these approaches to nuclear systems, where correlations are stronger and interactions more intricate. This also presents a distinct path from other approaches, such as a classical benchmark study that uses eigenvector continuation with the Lipkin-Meshkov-Glick model to evaluate GCM states \cite{extendedgcm_LMG, GCMLipkin}.

This paper is organized as follows: In Sec.~\ref{Theory}, we outline the theoretical framework and introduce the excitation and generator operators that form the basis of our adaptive ansatz. Section~\ref{Methodology} describes the procedure for mapping the nuclear Hamiltonian onto qubit space and outlines the structure of the quantum circuits employed in our simulations and the adaptive optimization strategy that illustrates how the Gray Code (GC) encoding formalism enhances the expressibility and efficiency of the ansatz construction, over the Jordan-Wigner (JW) transformation of One-Hot (OH) encoding. Section~\ref{Results} presents and analyzes our results for representative nuclear systems, highlighting the accuracy and scalability of the proposed approach. Finally, Sec.~\ref{Conclusion} summarizes the main findings and discusses potential extensions of this framework towards larger and more complex nuclear configurations.

\section{Theory}
\label{Theory}

The GCM is a powerful approach that links single-particle motion with collective behavior in many-body systems by constructing states as mixtures of wavefunctions labeled by collective variables that capture key correlations such as deformation, vibration, or pairing, and whose mixing yields access to both ground and excited states through the Hill–Wheeler Eq. (\ref{Hill-Wheeler integral equation}). These coordinates capture highly non-trivial correlations, including collective dynamics and shape mixing. The method relies on the principle that the wavefunction obtained from the energy eigenvalue equation in an orthonormal basis is consistent with the generalized eigenvalue equation in a non-orthonormal basis. The generalized eigenvalue equation in the context of GCM is known as the Green-Hill-Wheeler, or simply the Hill-Wheeler, equation.
We express the total wave function of a many-body system as a superposition of non-orthogonal basis~\cite {RingSchuckBook}
\begin{equation}
\label{eqn:conswf}
\Psi = \int g(q) \Phi(q)dq,
\end{equation}
where $\Phi(q)$ are the basis wave functions and $g(q)$ are the corresponding weights. The energy eigenvalue equation
\begin{equation}
    \label{energy_equation}
    H \Psi= E \Psi,
\end{equation}
takes the form of a generalized eigenvalue problem 
\begin{equation}
\label{Hill-Wheeler integral equation}
\int \left[ \mathcal{H}(q, q') - E \mathcal{N}(q, q') \right] g(q') \, dq' = 0,
\end{equation}
where $\mathcal{H}(q, q')$ and $\mathcal{N}(q, q')$ are the Hamiltonian and norm overlaps of generator states. These matrix elements are computed with the help of basis states $\phi(q)$ and $\phi(q')$ as,
\begin{align}
    \label{overlaps_equation}
    \mathcal{H}(q, q') = \braket{\Phi(q)|H|\Phi(q')}\notag \\\ \text{ and } \quad \mathcal{N}(q, q') = \braket{\Phi(q)|\Phi(q')}. 
\end{align}
The weight functions $g(q)$ are obtained by solving the Eq.~\eqref{Hill-Wheeler integral equation}. For computational purposes, we discretize the wavefunction as  
\begin{equation}
\Psi = \sum_{q} g(q) \Phi(q),
\end{equation}
to solve
\begin{equation}
\label{HillWheelerEquation}
\mathcal{H} \psi = E\mathcal{N}\psi,
\end{equation}
which is the Hill-Wheeler integral equation given by Eq.~\eqref{Hill-Wheeler integral equation} in matrix form.

To represent a quantum system, usually with an atomic, molecular, or nuclear system, the number of particles must remain conserved, and the basis should have correspondence with the physical system. For this purpose, the HF basis is widely employed as the set of generator states, since HF solutions provide physically meaningful, optimized single-particle determinants that capture the dominant mean-field features of the system. These HF states can be efficiently computed and naturally represent different shapes, deformations, or pairing configurations by varying appropriate constraints, making them ideal as building blocks for GCM~\cite{THOULESS1960225}. Using HF states as generators allows the GCM to incorporate fluctuations and correlations around the mean-field in a systematic, computationally tractable manner. The HF method plays a vital role in the foundations of computational quantum chemistry and nuclear physics, providing the best possible single-determinant approximation to the many-body ground state. It is widely used to approximate the ground state of many-body quantum systems. It is based on the idea that the best possible single Slater determinant can represent the system, obtained by minimizing the expectation value of the Hamiltonian 
\begin{equation}
\label{eq:HFenergy}
E_{\text{HF}} = \braket{\Phi_{\text{HF}} | H | \Phi_{\text{HF}}}\\
\end{equation}
while enforcing the condition that $\ket{\Phi_\text{HF}}$ remains a single Slater determinant. The solution of the equations
\begin{equation}
\label{eq:HFeq}
{f}^{\text{HF}} \ket{\varphi_i} = \epsilon_i \ket{\varphi_i},
\end{equation}
determines the Hartree-Fock orbitals, where $ {f}^{\text{HF}}$ is the Fock operator. This operator contains the kinetic energy, the external potential, and mean-field terms that describe the average effect of all the other particles. By iterating Eq.~(\ref{eq:HFeq}), the orbitals become consistent with the mean field they generate.  The Hartree-Fock ground state is then written as a product of creation operators acting on the vacuum
\begin{equation}
\label{eq:HFstate}
\ket{\Phi_{\text{HF}}} = \prod_{i=1}^{N} a_i^\dagger \ket{0} 
\equiv a_1^\dagger a_2^\dagger \cdots a_N^\dagger \ket{0} ,
\end{equation}
where $a_i^\dagger$ creates a particle in the $i$-th HF orbital. Thus, the HF method gives a physically meaningful and computationally manageable reference state. However, HF often falls short in accurately capturing excited states, particularly those dominated by particle correlation or multi-reference character. 

To address this, the UCCSD ansatz extends the HF framework by incorporating excitation operators, single and double particle-hole excitations, which generate a richer manifold of states encompassing both ground and excited configurations~\cite{Bartlett2007}. In practice, excited states can be accessed by applying UCCSD excitation operators to the HF reference state, effectively creating variational trial states beyond the single-determinant. This approach allows the capture of dynamic correlation effects, which are essential for accurate excitation energies. Moreover, by varying the excitation amplitudes, the ansatz can represent a range of excited states within the manifold spanned by the UCCSD operators.

The UCCSD ansatz provides a systematic way to include correlations beyond the mean-field HF description. Its wavefunction is defined as  
\begin{equation}
\label{eq:UCCSD_wf}
\ket{\Psi_{\text{UCCSD}}} = e^{T - T^\dagger} \ket{\Phi_{\text{ref}}},
\end{equation}
where $\ket{\Phi_{\text{ref}}}$ is usually the HF reference determinant and $T = T_1 + T_2$ contains single and double excitation operators. These $T_i$ are the transfer operators that transfer particles from one energy level to another and can be single or double operators depending on the number of particles transferred or excited. The singles operator,  
\begin{equation}
\label{eq:T1_operator}
T_1 = \sum_{i \in \text{occ}} \sum_{a \in \text{virt}} t_i^a \, a_a^\dagger a_i ,
\end{equation}
promotes one particle from an occupied orbital $i$ to a virtual orbital $a$, while the doubles operator,  
\begin{equation}
\label{eq:T2_operator}
T_2 = \frac{1}{4} \sum_{i,j \in \text{occ}} \sum_{a,b \in \text{virt}} t_{ij}^{ab} \, a_a^\dagger a_b^\dagger a_j a_i ,
\end{equation}
creates two simultaneous excitations. Because of the unitary exponential form in Eq.~\eqref{eq:UCCSD_wf}, the wavefunction remains normalized automatically, which makes it naturally suited for quantum computing implementations where normalization is essential~\cite{izmaylov2019unitary,paldus1970stability,paldus1988clifford}. 

In the GCM framework, UCCSD excitations serve as generator transformations. Instead of restricting the generator space to constrained HF solutions, one can construct correlated generator states of the form  
\begin{equation}
\label{eq:GCM_UCCSD}
\ket{\Phi(q)} = e^{ {X}(q)} \ket{\Phi_0},
\end{equation}
where $ {X}(q)$ is a generalized operator built from linear combinations of single or double excitation transfer operators $T_i$'s with $q$-dependent weights. This allows the generator coordinate $q$ to span correlated configurations, systematically enriching the GCM basis and improving its ability to capture strong correlations.  

Within the GCM framework, constructing an effective basis demands a set of physically meaningful generator states that capture the essential features of the system. As a natural starting point, the HF method provides a computationally efficient mean-field reference by representing the many-body wavefunction as a single Slater determinant. This approximation captures the dominant single-particle structure and serves as the foundation upon which the GCM is built. However, since the HF approach inherently neglects particles' correlation, it falls short of achieving quantitative accuracy. To overcome this limitation, we employ the UCCSD ansatz, which systematically introduces particle correlation through an exponential operator of single and double excitations acting on the HF state. Its unitary nature makes it particularly suited to quantum implementation, enabling the generation of a family of correlated states that form a robust, physically meaningful GCM basis. When combined with the GCM, the HF-based states serve as reference points, while UCCSD-type excitation operators act as generator transformations, producing a flexible, parametrized subspace that includes excited states. Solving the resultant generalized eigenvalue problem yields both ground and excited-state energies systematically. This integration leverages the computational efficiency and physical interpretability of HF, combined with the correlation capabilities of UCCSD, to provide a robust description of excited states in complex quantum systems.  
By integrating the GCM with HF and UCCSD-generated states, we establish a comprehensive theoretical scheme capable of describing both mean-field and correlated effects within a unified formalism. While conceptually powerful, this framework must be expressed in an implementable form suitable for quantum computation. The following section develops this transition by outlining the methodological workflow for implementing the GCM framework on quantum hardware via appropriate mappings and circuit-construction strategies.

In this way, HF, along with UCCSD within GCM, in the OH encoding of JW transformation, not only provides a physically transparent picture, but also a quantum-computing-compatible framework that balances accuracy with feasibility. Thus, making it a natural choice for extending many-body methods onto NISQ devices.

An alternative representation of the generator states is developed using the GC framework, a binary numbering scheme in which consecutive states differ by a single bit~\cite{gray1953pulse}. Within this perspective, the generator states are no longer viewed merely as HF references but as correlated configurations generated through compact, structured operators. This formulation naturally connects to coupled-cluster theory, as the action of UCCSD-type excitations can be systematically expressed within the GC formalism. The theoretical significance of this approach lies in providing a structured and efficient representation of correlated generator states within the GCM, while its computational advantages, including reduced circuit complexity and resource cost, are discussed in the Methodology section.

\section{Methodology}
\label{Methodology}
To translate the theoretical framework introduced in Sec.~\ref{Theory} into a form executable on quantum hardware, it is necessary to map fermionic operators onto qubit representations that can be realized as quantum circuits. This section presents the computational workflow underlying our approach, beginning with the QuGCM and its adaptive variant, ADAPT-GCIM, followed by the discussion of fermion-to-qubit mappings. We compare the standard JW transformation with the more compact GC encoding to highlight their relative efficiency and circuit complexity, thereby establishing the complete computational pipeline for quantum simulation.

\subsection{QuGCM}
\label{QuGCM}
\begin{figure*}[th]
    \centering
    \includegraphics[width=1\linewidth]{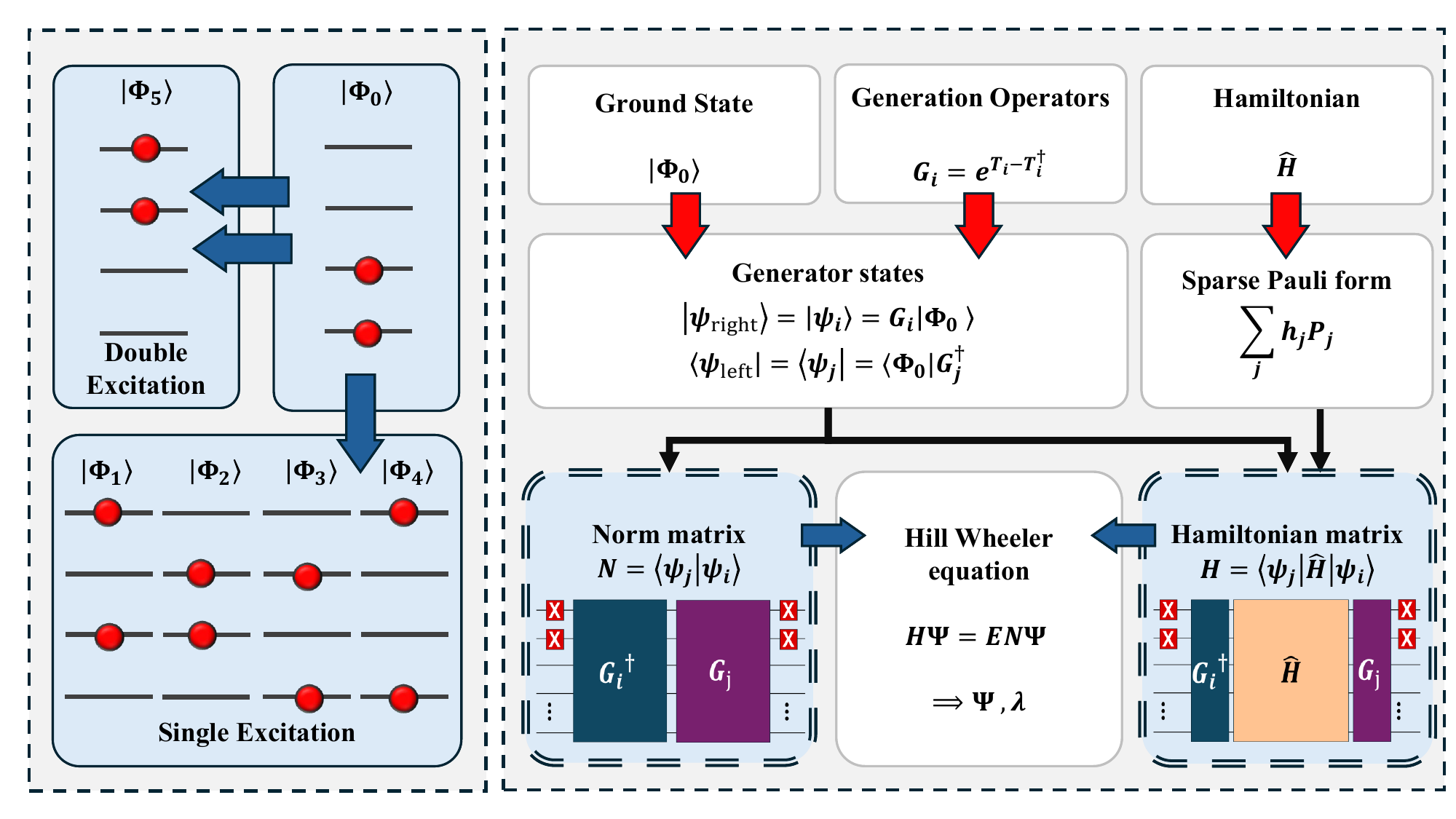}
    \caption{Flowchart of the hybrid QuGCM. A non-orthogonal basis is first constructed classically from a reference state. A quantum computer is then employed to calculate the Hamiltonian ($H$) and Norm ($N$) matrices in this basis. The system's energies ($E$) and wavefunctions ($\Psi$) are obtained by classically solving the resulting Hill-Wheeler generalized eigenvalue equation, $H \Psi = E N \Psi$.}
    \label{fiG:QuGCM_flowchart}
\end{figure*}

As discussed in Sec.~\ref{Theory}, the GCM requires solving the generalized eigenvalue problem, also known as the Hill-Wheeler equation given by Eq.~\eqref{HillWheelerEquation}. The GCM is adopted as a feasible method for the NISQ devices, referred to as QuGCM (Quantum GCM). The wave function $\psi$ together with the Hermitian Hamiltonian can be mapped into a qubit system using JW transformations of OH encoding~\cite{Batista2001, JordanWigner1928, bravyi2002fermionic, mcardle2020quantum}, and from this mapping, the Hamiltonian and the norm matrices are obtained, denoted by $H$ and $N$ respectively. Solving this equation provides the energy eigenvalues, usually expressed as a diagonal matrix $E = \text{diag}(\lambda_1, \lambda_2, \ldots)$, where the lowest $\lambda_i$ corresponds to the ground state energy of the system. This entire process can be performed using either classical or quantum computational methods, depending on the system size and available resources. The specific workflow for this hybrid approach, detailing the interaction between classical basis construction and quantum matrix calculation, is illustrated in Fig.~\ref{fiG:QuGCM_flowchart}. 

With an increase in the number of generator bases, the execution on a quantum computer becomes more relevant and useful.
To construct the generator states, we begin with a physical system and select its reference state $\ket{\phi_0}$ together with the corresponding Hamiltonian. Different possible excited states of the system can be obtained by applying excitation operators $T_i$ on this reference, transforming it into $\ket{\phi_i}$ as
\begin{equation}
\ket{\phi_i} = T_i \ket{\phi_0}\quad\text{ and} \quad \bra{\phi_j} = \bra{\phi_0} T_j^\dagger .
\end{equation}
The collection of all $T_i$ forms the set of generating functions or basis functions for the QuGCM. Following the framework described in Ref.~\cite{zheng2023quantum}, the anti-Hermitian combinations of these excitations are defined as $R_i = T_i - T_i^\dagger$, which preserve unitarity. Their action on the reference produces the correlated states
\begin{align}
R_i \ket{\phi_0} =\;&
\ket{\phi_i}, \quad R_i \ket{\phi_i} = -\ket{\phi_0}, \text{ and}\\
R_i (\ket{\phi_0} +\;&  \ket{\phi_i}) =\ket{\phi_i} - \ket{\phi_0}.
\end{align}
By exponentiating these $R_i$, one obtains unitary operators $G_i = e^{R_i}$ that generate correlated superpositions of the reference and excited configurations. These $G_i$ operators are convenient for quantum circuits since they are manifestly unitary. Their action defines the generator states
\begin{equation}
G_i \ket{\phi_0} = \ket{\psi_i}, \quad \bra{\psi_j} = \bra{\phi_0} G_j^\dagger .
\end{equation}
The first operator $G_0$ recovers the reference $\ket{\phi_0}$, while higher $G_i$ generate states that are superpositions of $\ket{\phi_0}$ and excited determinants $\ket{\phi_i}$. Extending this construction, one can use multiple generators simultaneously~\cite{fukutome1981group}. For instance, applying $G_i$ and $G_j$ successfully yields  
\begin{align}
\ket{\psi_{(i,j)}} &= G_i G_j \ket{\phi_0} \notag\\ &= \text{superposition of } \{ \ket{\phi_0}, \ket{\phi_i}, \ket{\phi_j},\ket{\phi_a}\},
\end{align}
where $\ket{\phi_a}$ represents additional states coupled through the two generators. Because the product of unitary operators is itself unitary, this procedure expands the accessible configuration space while maintaining normalization.

The matrix elements of the Hill-Wheeler equation are then computed from these generated states. Each element of the Hamiltonian and norm matrices is written as  
\begin{align}
H_{ij} &= \braket{\psi_j | H | \psi_i} = \braket{\phi_0 | G_j^\dagger H G_i | \phi_0}, \\
N_{ij} &= \braket{\psi_j | \psi_i} = \braket{\phi_0 | G_j^\dagger G_i | \phi_0}.
\end{align}

The matrices $H$ and $N$ are first calculated using quantum circuits and then supplied to the numerically solvable generalized eigenvalue problem $H \psi = E N \psi$ as shown in the final column of the Fig.~\ref{fiG:QuGCM_flowchart}. In this way, the integration of QuGCM with unitary excitations such as UCCSD provides a systematic framework for extracting ground and excited states of complex quantum systems using both classical and quantum resources.

The integration of the GCM with HF and UCCSD-generated states provides a comprehensive theoretical framework for constructing correlated many-body wavefunctions with high accuracy. While this formulation captures both mean-field and correlation effects within a unified scheme, it still depends on a predefined set of generator states, which may not always represent the most efficient or physically relevant basis for a given system. To overcome this limitation, an adaptive approach can be introduced to dynamically refine the basis construction. The next subsection presents the ADAPT-GCIM algorithm, an adaptive framework that iteratively builds the GCM basis by selecting the most influential generators, thereby enhancing both computational efficiency and physical fidelity.

\subsection{ADAPT-GCIM}
\label{ADAPT-GCIM}

\begin{figure*}[th]
    \centering
    \includegraphics[width=1\linewidth]{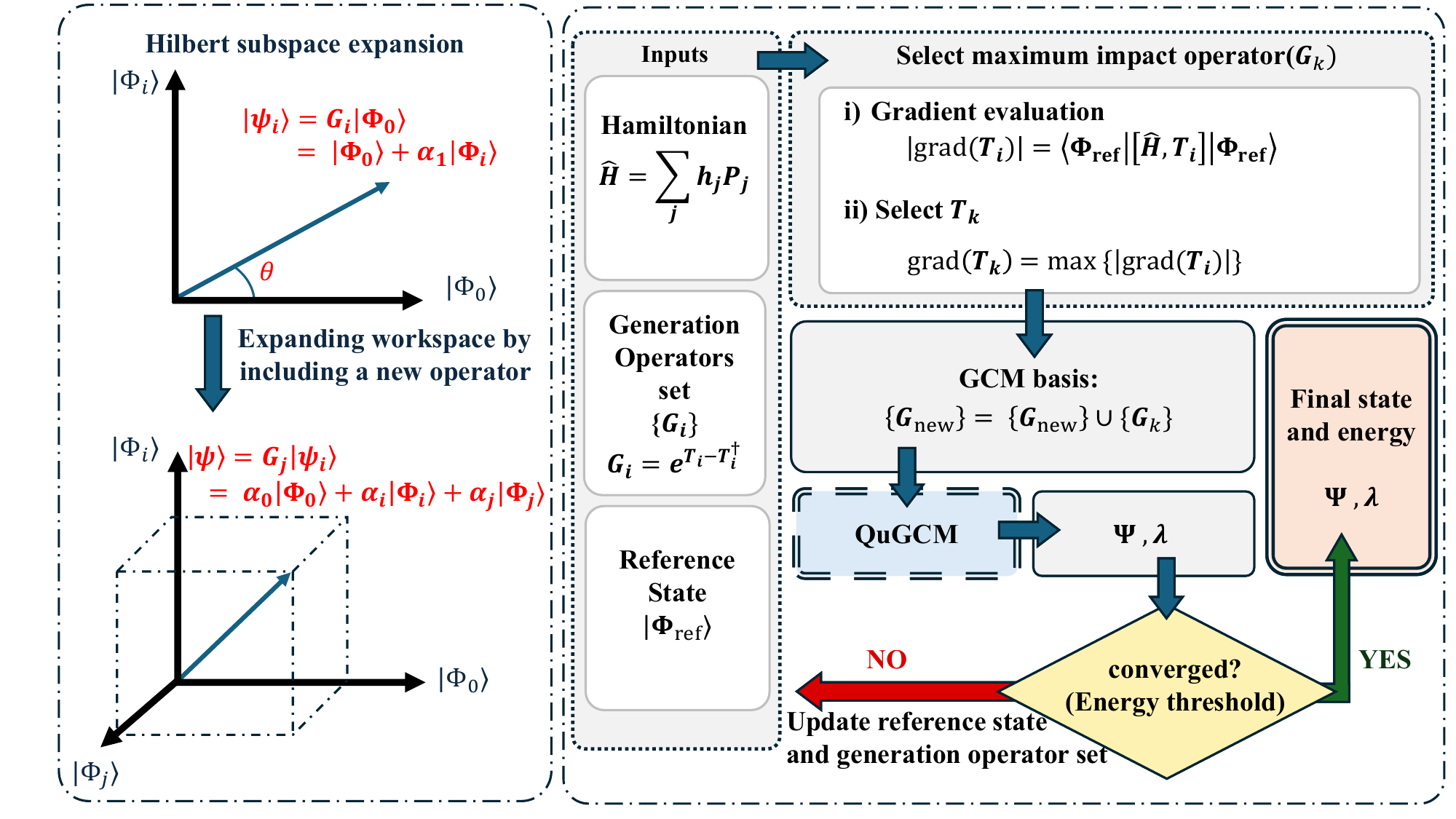}
    \caption{Flowchart for an adaptive quantum algorithm ADAPT-GCIM that iteratively refines the basis set for the QuGCM. The algorithm evaluates the impact of Generation Operators on the system's energy, selects the most influential one to add to the QuGCM basis, and then solves the GCM equations. This loop continues until the calculated energy converges, resulting in the final state and energy of the system. On the left, a conceptual diagram of the adaptive Hilbert space expansion. The ADAPT algorithm iteratively expands the Hilbert subspace, starting from an initial 2D basis defined by a reference state $|\Phi_0\rangle$ and a generated state $|\Phi_i\rangle$. In each step, a new basis vector $|\Phi_j\rangle$ is added by selecting the most impactful operator $G_j$, effectively increasing the subspace dimension (e.g., from 2D to 3D). This allows the wavefunction $|\Psi\rangle$ to be constructed efficiently by exploring only the most relevant portion of the Hilbert space.}
    \label{fig:adaptgcimflowchart}
\end{figure*}

Building upon the theoretical foundation established in Sec.~\ref{QuGCM}, this section introduces the ADAPT-GCIM, a dynamic extension of the GCM designed for efficient implementation on quantum hardware. Unlike conventional GCM, which relies on a fixed set of generator states, ADAPT-GCIM constructs the basis iteratively by evaluating the impact of each generator on the system's energy. At each step, the operator contributing the most to energy reduction is selected and incorporated into the basis, ensuring that the Hilbert space is expanded along the most relevant physical directions. This adaptive procedure enables systematic accuracy improvement while minimizing computational resources, making it particularly well-suited for near-term quantum devices.

This idea is closely related to adaptive variational methods such as ADAPT-VQE~\cite{grimsley2019adaptive} and is implemented here in the ADAPT-GCIM framework. In this approach, the generators are not fixed in advance but are dynamically chosen based on their contributions to the energy levels. The most impactful generator is identified by evaluating the energy gradient of each candidate operator with respect to the Hamiltonian, then selecting the one with the maximum gradient for evaluation on NISQ devices. Fig.~\ref{fig:adaptgcimflowchart} demonstrates the workflow of the ADAPT-GCIM process, where we choose the most impactful generator operator and use it to find the energy eigenvalue, consecutively, till the convergence of the obtained energies. This one-by-one selection of the impactful operator includes the expansion of the working Hilbert subspace, as shown in Fig.~\ref{fig:adaptgcimflowchart}. Consequently, the new state $| \Psi \rangle$ becomes a linear combination of a larger set of basis states, allowing for a more accurate representation of the system's true wavefunction. This process of expansion continues till the subset of basis contributing to the wavefunction is fully occupied. Thus, an adaptive method via successive iterations constructs the wavefunction, including the only portion of Hilbert space where the solution resides.

To implement these frameworks on quantum hardware, it is necessary to translate the system's fermionic Hamiltonian into qubit operators that can be directly implemented on a quantum circuit. This translation is achieved through fermion-to-qubit mappings, the most standard of which is the JW transformation with OH encoding, while other, more compact schemes, such as GC encoding, offer advantages in reducing circuit depth. Both methods allow us to represent the Hamiltonian and the generator operators in terms of Pauli matrices, the native language of quantum computers. The next sub-sections outline this process, comparing the standard JW transformation with the more compact and resource-efficient GC encoding.

\subsection{JW transformation}
We begin with the generic second-quantized Hamiltonian for a fermionic system, expressed as
\begin{equation}
H = \sum_{ij} h_{ij} a_i^\dagger a_j + \tfrac{1}{2}\sum_{ijkl} h_{ijkl} a_i^\dagger a_j^\dagger a_k a_l ,
\end{equation}
where $a_i^\dagger$ and $a_j$ are fermionic creation and annihilation operators, $h_{ij}$ represent one-body terms (kinetic and external potential contributions), and $h_{ijkl}$ encode two-body interactions. To solve the Hill-Wheeler equation within the GCM, matrix elements of this Hamiltonian must be computed between generator states. However, on a quantum computer, the fermionic operators must first be mapped to tensor products of Pauli operators acting on qubits.

The JW transformation provides a straightforward and systematic mapping by ordering the fermionic orbitals along a line and associating each orbital with a qubit~\cite{ JordanWigner1928}. The ladder operators take the form
\begin{equation}
a_j^\dagger = \tfrac{1}{2}(X_j - i Y_j)\prod_{k<j} Z_k , 
\quad 
a_j = \tfrac{1}{2}(X_j + i Y_j)\prod_{k<j} Z_k ,
\end{equation}
where $X_j$, $Y_j$, and $Z_j$ are the Pauli matrices acting on the $j$-th qubit. The string of $Z$ operators enforces the correct antisymmetry under particle exchange, guaranteeing that fermionic statistics are preserved. For example, consider a system with two fermionic orbitals. The number operator $n_0 = a_0^\dagger a_0$ is mapped as
\begin{equation}
n_0 = \tfrac{1}{2}(I - Z_0),
\end{equation}
while the hopping operator $a_0^\dagger a_1 + a_1^\dagger a_0$ becomes
\begin{equation}
a_0^\dagger a_1 + a_1^\dagger a_0 = \tfrac{1}{2}(X_0 X_1 + Y_0 Y_1).
\label{eqn:eq24}
\end{equation}

To illustrate this concept, we consider a model system consisting of two fermions distributed across four spin–orbitals. This minimal system serves as a standard benchmark because it represents the simplest nontrivial case that exhibits both single- and double-particle–hole excitations. In this setup, the Hartree–Fock ground state corresponds to the two lowest-energy orbitals being occupied, denoted in the occupation number basis as $\ket{1100}$. Excited configurations such as the singly excited $\ket{1010}$ and the doubly excited $\ket{0011}$ can be defined in an analogous manner. The occupation number basis maps naturally onto a qubit register, where each qubit encodes the occupancy of a corresponding orbital, with $\ket{1}$ indicating an occupied state. 
This mapping ensures a one-to-one correspondence between the fermionic Fock space and the qubit Hilbert space, allowing any fermionic transfer operator or its combinations can be systematically expressed as a sum of Pauli strings, as written in Eq.~\eqref{eqn:eq24}. This illustrates that any fermionic operator can be systematically expressed as a sum of Pauli strings, with a natural correspondence to the occupied position by $1$ and the unoccupied by $0$. The state of qubits shows a particle is present whenever the qubit is $1$, or otherwise stated. This creates an orthonormal basis for the wavefunction in a quantum computer. However, a limitation arises because the Pauli strings grow linearly in length with the orbital index, meaning that operators acting on high-index orbitals involve long chains of $Z$’s, increasing the circuit depth on quantum hardware. This limitation can be overcome by those specific configuration methods that conserve the particles in a system.

The UCCSD method provides an efficient ansatz for capturing nucleon correlations. When combined with the JW  transformation, the fermionic excitation operators are directly mapped into qubit operators using strings of Pauli matrices. The exponential structure of the UCCSD ansatz can then be written as
\begin{equation}
|\psi(\theta)\rangle = e^{T(\theta) - T^\dagger(\theta)} | \phi_0 \rangle,
\end{equation}
where $T(\theta)$ is the transfer operator which contains single and double excitations. Under the JW transformation, these excitations translate into products of $X$, $Y$, and $Z$ operators acting on qubits, making UCCSD suitable for implementations in quantum algorithms. For example, the transfer operator $T$ that maps the state $\ket{1100}$ to $\ket{1001}$, $\ket{0110}$, and $\ket{0011}$ can be expressed in terms of creation and annihilation operators. The corresponding excitation operators that act on $\ket{1100}$ are  
\begin{align}
&T_{\ket{1100} \rightarrow \ket{1001}} = a_3^\dagger a_1 ,\quad
T_{\ket{1100} \rightarrow \ket{0110}} = a_2^\dagger a_0 \quad
\notag \\ \text{ and }\quad 
&T_{\ket{1100} \rightarrow \ket{0011}} = a_2^\dagger a_3^\dagger a_0 a_1 .
\end{align}
where $a_i^\dagger$ and $a_i$ are the fermionic creation and annihilation operators acting on orbital $i$. From these excitation operators, an anti-Hermitian generator $R(\theta)$ is constructed~\cite{k-UpCCGSD-paper,thouless2014quantum}. The corresponding unitary operator for the UCC method is then obtained by exponentiation as
\begin{equation}
    G(\theta) = e^{R(\theta)} = e^{T(\theta) - T^\dagger(\theta)},
\end{equation}
where $\theta$ is a variational parameter that controls the relative contribution of each excitation operator. Applying this generator operator to the ground state produces the UCC ansatz. However, for the generalized eigenvalue equation (here the Hill Wheeler Eq.~(\ref{HillWheelerEquation})), the resulting wavefunction for a non-orthogonal basis remains independent of the parameters and relies only on matrix inversion and multiplication to obtain the respective coefficients or weights associated with them. Therefore, considering the parameter to be any constant while constructing individual generation operators $G_i$ for the generator basis $\psi_i$, will result in the solution of the Hill Wheeler equation, unless the norm matrix becomes singular.

In this framework, we parameterize the excitation operators corresponding to single and double excitations. For single excitations from occupied orbital $i$ to virtual orbital $a$, the parameterized operator is  
\begin{align}
    T_{ai}(\theta_{ai}) &= \theta_{ai}\, a_a^\dagger a_i ,
\end{align}
where $\theta_{ai}$ is the variational parameter associated with this excitation. For double excitations from $(i,j)$ to $(a,b)$, the operator is written as  
\begin{align}
    T_{abij}(\theta_{ai}, \theta_{bj}) &= \theta_{ai}\, a_a^\dagger a_i 
    + \theta_{bj}\, a_b^\dagger a_j ,
\end{align}
where $\theta_{ai}$ and $\theta_{bj}$ control the contributions of each excitation operators.  
In general, we denote an excitation operator as $T_i$, where the index $i$ 
labels either a single- or double-excitation operator. Each such operator 
generates a unitary transformation of the form  
\begin{equation}
    G_i(\theta_i)  = e^{R_i(\theta_i)}= e^{\theta_i(T_i - T_i^\dagger)},
\end{equation}
which ensures their implementations as quantum gates. 

As demonstrated in Appendix~\ref{appendix_parameter_indpendence}, the results obtained from the QuGCM are independent of the parameter values, provided the generator forms a set of distinct basis elements. Consequently, the parameters may be fixed to any constant value, with the exception of $0$ or multiples of $2\pi$. These exclusions are necessary to ensure that the operator $\mathrm{G}_i$ does not reduce to the identity. Taking this parameter or $\theta$ to be $\pi/4$ constant for the construction of each generation operator as UCC operators for singles as well as doubles excitations, to get different excited basis. Thus making generator operators as
\begin{equation}
    G_i = e^{\frac{\pi}{4} (T_i-T_i^\dagger)}.
\end{equation}
The values of $G_i$ become unity for $\theta$ being an integral multiple of $\pi$, making the reference basis as all other generator bases, resulting in an all-ones Norm matrix which is singular. The inverse of such a singular matrix is not possible and thus yields spurious values in the QuGCM or ADAPT-GCIM.
\begin{equation}
    G_i = e^{n\pi (T_i-T_i^\dagger)} = I ,\quad n \in \mathbb{Z}.
\end{equation}
In practice, these $G_i$ obtained as exponentials of $R_i$ are implemented on quantum hardware using Trotterization techniques. One is the Suzuki-Trotter expansion of order 2, achieved via the PauliEvolutionGate operation in Qiskit~\cite{Qiskit}.

\subsection{GC encoding}
\label{GC_encoding}
The effectiveness of the QuGCM strongly depends on the chosen fermionic-to-qubit encoding. This choice determines the complexity of the operators that run on the quantum computer. Therefore, it is clear to use GC encoding, proven to be more efficient and less vulnerable to noise than the JW transformation, owing to quantum circuits with fewer qubits, fewer quantum gates, and shallower circuits when the ansatz and quantum circuits are constructed from the same principle using similar operators. For this, we compare the QuGCM method implementations using JW transformation with a more efficient GC encoding. We use a model of two fermions in four spin-orbitals and their excitation operators to show the clear advantages of the GC encoding~\cite{MatteoGC, Siwach2021, singh2025advancingquantumsimulationsnuclear}.

Our system's physical Hilbert space is six-dimensional. We define a basis using six Slater determinants, $|d_i\rangle$, which are ordered by the indices of the occupied orbitals. Using a qubit ordering of $|q_0 q_1 q_2 q_3\rangle$ for orbitals $0, 1, 2, 3$, the basis is defined by slater determinant $\ket{d_i}$. The JW encoding uses four qubits to represent these states directly. Because the JW representation is a direct occupation-number encoding, each Slater determinant maps to a unique computational basis bitstring without any further transformation. This is why the configuration in column 2 (Slater determinant) and the JW bitstring in column 3 appear identical in Table~\ref{tab:state_mapping_final}. The GC encoding, however, maps this six-dimensional space into a more compact three-qubit register, $|g_i\rangle$. The complete mapping is shown in Table~\ref{tab:state_mapping_final}.

\begin{table*}[t]
\centering
\large 
\caption{Encoding of $4$-particle Slater determinants $|d_i\rangle$ into qubit states using Jordan-Wigner (JW) and Gray Code (GC) mappings. Direct one-to-one maps from Slater to JW state using $4$ qubits, which is the same as the number of particles. It is compressed to a smaller number, $3$ qubits for the GC state.}
\label{tab:state_mapping_final}
\begin{tabular}{|c|c|c|c|}
\hline
\textbf{State Index} & \textbf{Slater determinant} & \textbf{JW State} & \textbf{GC State} \\
\textbf{ ($\mathbf{i}$)} & \textbf{  ($\mathbf{|d_i\rangle}$)} & \textbf{ ($\mathbf{4}$ Qubits)} & \textbf{($\mathbf{|g_i\rangle}$)} \\\hline
1 & $|1100\rangle$ & $|1100\rangle$ & $|000\rangle$ \\ \hline
2 & $|1010\rangle$ & $|1010\rangle$ & $|001\rangle$ \\ \hline
3 & $|1001\rangle$ & $|1001\rangle$ & $|011\rangle$ \\ \hline
4 & $|0110\rangle$ & $|0110\rangle$ & $|010\rangle$ \\ \hline
5 & $|0101\rangle$ & $|0101\rangle$ & $|110\rangle$ \\ \hline
6 & $|0011\rangle$ & $|0011\rangle$ & $|111\rangle$ \\ \hline

\end{tabular}
\end{table*}

\begin{figure}[hbt!]
    \centering
    \includegraphics[width=\linewidth]{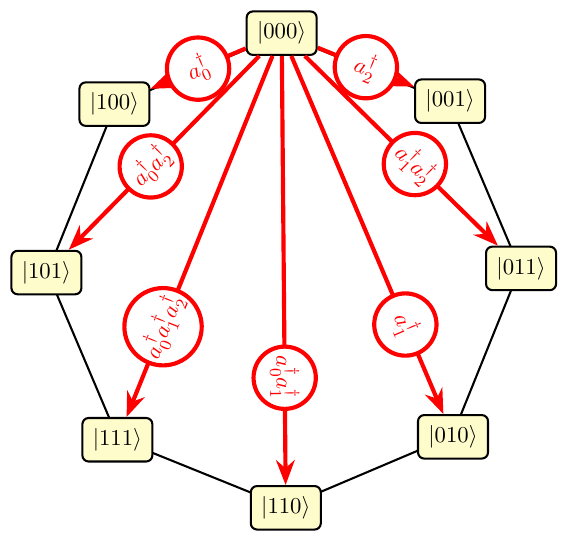}
    \caption{Schematic diagram showing all conversions from a reference in GC encoding.}
    \label{fig:GC_ref_state_eg000}
\end{figure}

This organizes the occupation basis states in binary-reflected Gray order. In this scheme, consecutive basis states differ by only a single qubit flip, thereby reducing the number of operations required to connect them. The QuGCM framework constructs its basis from a set of unitary generators $G_k = e^{R_k}$, which are derived from anti-hermitian operators $R_k = T_k - T_k^\dagger$. Each $T_k$ represents a fermionic excitation that transitions between basis states. While the JW transformation is standard, the GC encoding offers a significant computational advantage by using these fermionic excitations. The cyclic structure of the GC basis is built on successive single-bit flips, a process that can be implemented on qubits using creation and annihilation operators. This property ensures that any state is reachable from any other through combinations of these fundamental fermionic operations. These fermionic creation or annihilation operations create $T_K$ operators for GC also, in the same manner as that of JW. Crucially for QuGCM, this allows the GC encoding to replicate the behavior of general UCCSD operators. By applying these $T_k$ operators to a reference state, one can generate a superposition with other mapped states. For example, the reference GC state $\ket{000}$ is excited to other states as shown in Fig.~\ref{fig:GC_ref_state_eg000}. This creates the non-orthogonal generator basis required by the method. This approach effectively mirrors the role of UCC operators but with compressed and hardware-friendly implementations. Decomposing key excitations into the sum and product of Pauli matrices is simpler and requires fewer quantum gates in the GC encoding than the JW transformation.

For the single excitation $|d_1\rangle \rightarrow |d_4\rangle$, one can construct an operator and simplify to get it into the form of Pauli matrices. This transition moves a fermion from orbital 0 to 2, while the fermion at orbital 1 is unchanged. The operator is $T_{\ket{d_1}\to\ket{d_4}} = a_2^\dagger a_0$. For the JW encoding, the anti-hermitian generator $R_{\ket{d_1}\to\ket{d_4}}^{\text{JW}} = a_2^\dagger a_0 - a_0^\dagger a_2$ requires long parity-tracking $Z$ strings, resulting in a non-local operator:
\begin{equation}
R_{\ket{d_1}\to\ket{d_4}}^{\text{JW}} = \frac{i}{2} ( X_2 Z_1 Y_0 - Y_2 Z_1 X_0).
\end{equation}
This operator acts on three different qubits ($q_0, q_1, q_2$), making it complex to implement on hardware. For the GC encoding, the same physical process is a simple transition between compact states $|g_1\rangle=|000\rangle$ and $|g_4\rangle=|010\rangle$. This single bit-flip $a_1^\dagger$ is generated by a much simpler, local operator acting only on the second qubit ($q_1$):
\begin{equation}
R_{\ket{g_1}\to\ket{g_4}}^{\text{GC}} = - i Y_1 Z_0.
\end{equation}
The benefit of GC encoding is clear, as a 3-qubit non-local JW operator becomes a 2-qubit local GC operator. This greatly simplifies the corresponding quantum circuit, reducing gate counts and errors.

Consider the case of double excitation $|d_1\rangle \rightarrow |d_6\rangle$. This transition moves both fermions from orbitals $\{0,1\}$ to $\{2,3\}$, driven by the operator $T_{\ket{d_1}\to\ket{d_6}} = a_3^\dagger a_2^\dagger a_1 a_0$. In the JW encoding, this operator is highly non-local. The generator $R_{\ket{d_1}\to\ket{d_6}}^{\text{JW}}$ expands into a sum of eight distinct Pauli strings that each act on all four qubits:
\begin{align}
R_{\ket{d_1}\to\ket{d_6}}^{\text{JW}} = -\frac{i}{8} & (X_3 X_2 X_1 X_0 - Y_3 X_2 Y_1 X_0 \nonumber \\
                                &- X_3 Y_2 X_1 Y_0 + Y_3 Y_2 X_1 X_0 \nonumber \\
                                &+X_3 X_2 Y_1 Y_0 - Y_3 X_2 X_1 Y_0 \nonumber \\
                                &- X_3 Y_2 Y_1 X_0 + Y_3 Y_2 Y_1 Y_0).
\end{align}
Exponentiating this operator requires a very deep and complex quantum circuit. While in GC encoding, this process corresponds to the transition $|g_1\rangle=|000\rangle \rightarrow |g_6\rangle=|111\rangle$. While this flips all three logical bits, the resulting generator is significantly simpler than its JW counterpart:
\begin{equation}
R_{\ket{g_1}\to\ket{g_6}}^{\text{GC}} = \frac{i}{4} (X_2 Y_1 X_0 + X_2 X_1 Y_0 - Y_2 X_1 X_0 - Y_2 Y_1 Y_0).
\end{equation}
This representation of the GC generator contains only four terms, compared to the eight terms in the JW representation. This consistent reduction in operator complexity for both single and double excitations makes the GC encoding a superior choice for implementing the QuGCM on quantum computers.

\section{Results} \label{Results}

In this section, we present the outcomes obtained from the QuGCM and the ADAPT-GCIM approach. These are compared against the performance of pre-existing methods, including standard VQE, ADAPT-VQE, and VQD. The analysis has been carried out for three representative systems, highlighting both the accuracy and efficiency of the schemes under different encoding strategies. 
We first examine the deuteron system~\cite{Siwach2021}, which has a relatively simple Hamiltonian with low correlation energy, to validate the developed QuGCM and ADAPT-GCIM methods described in Sec. \ref{Methodology}. It's Hilbert space is spanned by only two occupation-number configurations, requiring just two qubits in the JW representation and providing a straightforward benchmark for testing the methods. The system $^{38}$Ar represents an intermediate-scale problem in our study, where the shell-model truncation generates a Hilbert space of moderate dimensionality. The many-body basis constructed from the relevant Slater determinants maps to a six-qubit representation in the JW encoding, providing a configuration space that is sufficiently rich to exhibit meaningful correlation effects while remaining tractable for simulations under both JW and GC encoding strategies, for both QuGCM and ADAPT-GCIM frameworks. Finally, for the larger and more complex system $^{6}$Li, we employ a reduced number of qubits using the GC encoding, which enables an efficient and practical simulation while maintaining accuracy~\cite{singh2025advancingquantumsimulationsnuclear}. Although $^{6}$Li contains fewer nucleons, the effective model space used to describe its correlations involves more active orbitals and richer configuration mixing than in $^{38}$Ar, leading to a significantly larger underlying Hilbert space composed of the twelve qubits in the direct JW representation and making it a more demanding system for quantum simulation.

To ensure a fair comparison of evaluation among the different methods, we adopt a standard simulation across all three systems. Noiseless results are obtained using a state-vector simulator to determine the fundamental and theoretical limits of the QuGCM and ADAPT-GCIM frameworks. For realistic performance assessment, noisy simulations are performed using the FakeBrisbane simulator~\cite{FakeBrisbane}, which mimics the noise profile of IBM’s Brisbane device and provides results as if the computation is done on a real Brisbane device. Each noisy energy estimate is derived from 100 independent runs, each with 8,192 shots, using the median absolute deviation (MAD) as the primary measure of statistical uncertainty. The number of runs and shots ensures that computational resources are used efficiently, with manageable noise and MAD errors. In all the cases, results are benchmarked against exact results obtained via classical diagonalization and established VQE-based methods.
\subsection{Deuteron system}
\begin{figure}[t]
    \centering
    \includegraphics[width=0.8\linewidth]{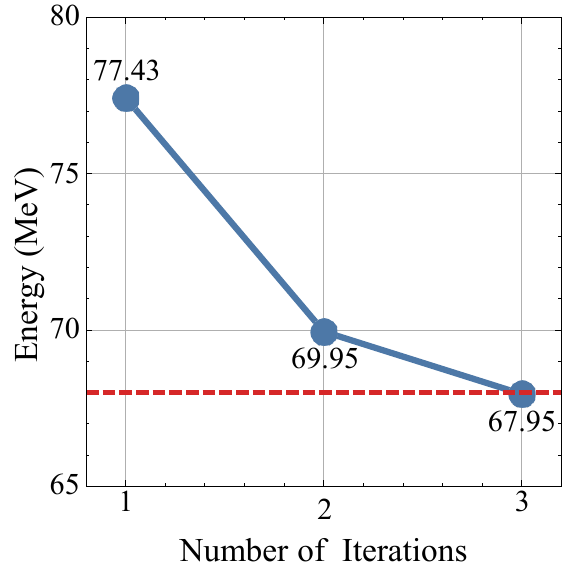}
    \caption{Convergence of the deuteron ground-state energy with ADAPT-GCIM iterations. The method achieves rapid convergence to the expected ground-state value (shown as the dashed line) within three iterations.}
    \label{fig:convergence}
\end{figure}

For the deuteron, the wave function has a mixture of $L=0$ (S-wave) and $L=2$ (D-wave), so the potential must include a tensor term. We use the Reid68 soft-core potential with a central term $V_C(r)$, tensor term $V_T(r) S_{12}$, and spin-orbit term $V_{LS}(r) \vec{L}\cdot \vec{S}$~\cite{reid1968local}. Matrix elements of $S_{12}$ and $\vec{L}\cdot \vec{S}$ are used to build the Hamiltonian in the single-particle basis. For quantum simulation, each $n$ and $l$ combination is assigned a qubit. For $N=1$ and $l=0,2$, the JW Hamiltonian is
\begin{align}
H &= 116.5294\,I - 78.85401\,Z_0 - 37.67542\,Z_1 \nonumber \\
&- 12.88914\,(X_0 X_1 + Y_0 Y_1),
\end{align}
which is real and Hermitian. Off-diagonal terms naturally include the S-D mixing. We use just two qubits for our test purpose of a smaller system; however, including more qubits to include higher orbital excitations yields more accurate energy eigenvalues~\cite{Siwach2021}. We applied QuGCM and ADAPT-GCIM to get energy eigenvalues. Table \ref{tab:tablefordeuteron} shows the comparison between the reference VQE energies and our QuGCM results. The energies reproduce the reference VQE results accurately while using no parameters and the fewest circuit elements, showing that QuGCM is resource-efficient without losing precision.
\begin{table*}[t]
    \centering
    \caption{Ground-state energies of the deuteron system computed using various quantum computational methods with the Reid68 potential. The true value obtained by the diagonalization is $67.948$ MeV \cite{Siwach2021}.}
    \begin{tabular}{|l|c|c|c|c|}
        \hline
        \textbf{Method} & {\textbf{QuGCM}} & {\textbf{ADAPT-GCIM}} & {\textbf{VQE}} & {\textbf{ADAPT-VQE}} \\ \hline
        without noise & 67.9476 & 67.9476 & 67.9476(0) & 67.9476(0) \\ \hline
        with noise & 67.94871(111) & 67.94786(26) & 67.94917(138) & 67.94813(34) \\ \hline
    \end{tabular}
    \label{tab:tablefordeuteron}
\end{table*}
Since ADAPT-GCIM selects operators iteratively based on energy gradients until convergence. Thus, we have calculated the convergence of the ground-state energy as a function of increasing ADAPT-GCIM iterations and plotted the results in Fig.~\ref{fig:convergence}. Starting from an initial trial state with higher energy, the method converges rapidly to the expected ground-state value within three iterations, demonstrating both efficiency and stability of operator selection in the adaptive scheme. Importantly, arbitrary choices of variational parameter $\theta$ in the exponential operators $T-T^\dagger$ lead to the same converged energy, confirming the theta independence for the same value of theta in all the ansatz constructions. The plot showing independence of the parameter for the simplest case, where a single parameter controls the admixture of two states, is presented in Appendix~\ref{appendix_parameter_indpendence}.

\begin{figure}
    \centering
    \includegraphics[width=\linewidth]{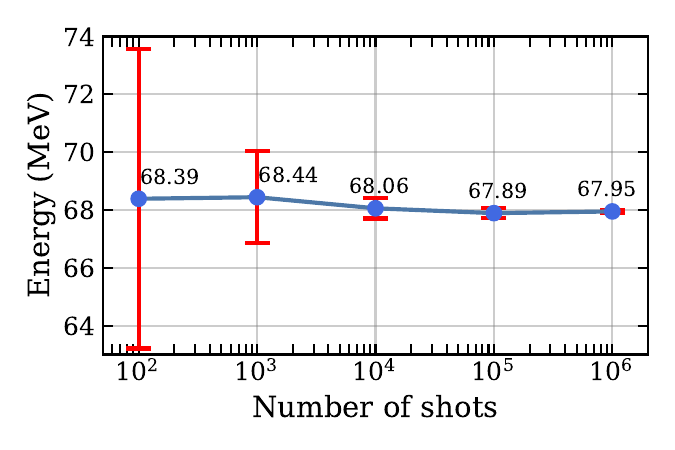}
    \caption{Ground-state energy estimates for the deuteron system as a function of the number of measurement shots. Statistical fluctuations at low shot counts decrease as the number of shots increases, with results converging toward the exact value.}
    \label{fig:shots}
\end{figure}

Figure~\ref{fig:shots} shows the energy estimates as a function of the number of measurement shots. At low shot counts, the statistical uncertainty is large because the measurement can't sample the wave function completely. As the number of shots increases, the error bars shrink, and the results stabilize around the exact value. This demonstrates that the proposed QuGCM and ADAPT-GCIM methods yield comparable results under the same circumstances of shot requirements when compared to variation-based VQE and ADAPT-VQE methods.

The deuteron case validates these methods under simple conditions for the JW mapping with a smaller number of gates and circuit runs. This indicates that for small systems such as the deuteron, the GCM-based method achieves performance comparable to variational methods while using a hardware-friendly operator structure.

We now extend the analysis to the argon system, where the Hilbert space is significantly larger, highlighting the differences between JW and GC. In the JW mapping, the Hamiltonian representation requires multiple Pauli strings, and the UCC operators generate excitations through standard creation and annihilation operators. Using GC encoding, the same set of excitations can be mapped to a compressed operator structure as discussed in Sec .~\ref{GC_encoding}, effectively reducing circuit depth. The resource gain is evident and discussed in detail in the section below, as fewer entangling gates are required to prepare the ground state, while maintaining the same level of accuracy.  

\subsection{\texorpdfstring{$^{38}$Ar}{38Ar}}
For the $^{38}$Ar nucleus, modeled as a $^{28}$Si core with two holes in the $sd$-shell, the valence space consists of the $s_{1/2}$ and $d_{3/2}$ orbitals using the USDB interaction~\cite{RichterUSDA_USDB,singh2025advancingquantumsimulationsnuclear}. In the $m$-scheme, this requires $N=6$ qubits, while GC encoding allows further compression depending on direct mapping to different qubits. Both encoding schemes are applied to the two classes of methods considered here: the variational family (VQE and its adaptive extensions) and the generator coordinate method (GCM and ADAPT-GCIM). For both GC and JW, the qubit form of the basis states and individual terms of the Hamiltonian formed for the $0^+$ state using the sum of sparse Pauli matrices is given in Appendix~\ref{app:Ar38}.

Since VQE and its adaptive variant rely on a variational ansatz with trainable parameters, the resulting energies can vary slightly between runs, even in noiseless simulations. By contrast, QuGCM and ADAPT-GCIM yield exact deterministic results in the noiseless case, independent of repetitions. To provide a fair comparison, we report values from both approaches directly as obtained, using the exact diagonalization results as a common reference. Both noiseless and noisy scenarios are explored under JW and GC mappings, and values are listed in Tables~\ref{tab:energy_comparison_Ar38_JW} and \ref{tab:energy_comparison_GC_Ar38}.

For the $^{38}$Ar under JW transformation, the energy spectra of the lowest $0^+$, $1^+$, and $2^+$ states are calculated. The accuracy of the eigenvalues demonstrates that ADAPT-GCIM and QuGCM achieve a greater energy reduction with fewer quantum gates than standard VQE or ADAPT-VQE.  Excited states such as the second $0^+$ and $2^+$, named $0^+_2$ and $2^+_2$, are accessed using the VQD algorithm~\cite{Higgott2019variationalquantum}, where QuGCM again shows consistent accuracy. Subscripts are used when there are two or more energy levels, including excited states, corresponding to the same $J^+$ value. The corresponding energies obtained are summarized in  Table \ref{tab:energy_comparison_Ar38_JW} and the values are compared against those obtained from VQE and ADAPT-VQE reported in the literature~\cite{singh2025advancingquantumsimulationsnuclear}.

\begin{table*}[t]
\centering
\caption{Comparison of energy eigenvalues for different $J^\pi$ states of $^{38}$Ar using VQE, ADAPT-VQE~\cite{singh2025advancingquantumsimulationsnuclear}, and calculated using QuGCM and ADAPT-GCIM for JW, with and without noise. Exact diagonalization values are shown for reference.}
\label{tab:energy_comparison_Ar38_JW}
\begin{tabular}{|c|c|c|c|c|c|c|}
\hline
\textbf{$\mathbf{J^\pi}$} & {\textbf{Exact}} & \textbf{Noise} & {\textbf{VQE}} & {\textbf{ADAPT}} & {\textbf{QuGCM}} & {\textbf{ADAPT}} \\
 & & \textbf{level} & & {\textbf{VQE}} & & {\textbf{GCIM}} \\
\hline
\multirow{2}{*}{$0_1^+$} & -152.677 & Without Noise & -151.7937(4477) & -152.3999(8798) & -152.677(0) & -152.677(0) \\
\cline{3-7}
 & & With Noise & -139.8726(94223) & -141.6124(72305) & -155.153(2476) & -155.629(2952) \\
\hline
\multirow{2}{*}{$2_1^+$} & -151.093 & Without Noise & -151.0934(18) & -150.8583(11249) & -151.093(0) & -151.093(0) \\
\cline{3-7}
 & & With Noise & -149.6657(5218) & -142.1343(45646) & -145.798(5295) & -145.967(5126) \\
\hline
\multirow{2}{*}{$2_2^+$} & -149.700 & Without Noise & -149.7003(0) & -149.7034(1) & -149.7003(0) & -149.7003(0) \\
\cline{3-7}
 & & With Noise & -149.7002(2) & -149.5180(5566) & -148.515(1185) & -148.185(1515) \\
\hline
\multirow{2}{*}{$0_2^+$} & -149.225 & Without Noise & -137.4674(187408) & -140.6414(144897) & -149.225(0) & -149.225(0) \\
\cline{3-7}
 & & With Noise & -110.8731(305988) & -129.7738(210019) & -150.959(1734) & -149.638(411) \\
\hline
\multirow{2}{*}{$1^+$} & -149.104 & Without Noise & -148.8602(6486) & -148.5788(42723) & -149.2677(1637) & -149.2677(1637) \\
\cline{3-7}
 & & With Noise & -136.1554(45441) & -141.8163(20340) & -144.263(4841) & -144.207(4897) \\
\hline
\end{tabular}
\end{table*}

Under GC encoding, the number of required qubits is significantly reduced compared to the JW scheme: three for the $0^+$ state, two for the $1^+$ state, and only one for the $2^+$ state. The implementation costs, approximated by the number of quantum gates, are used for fermionic operators in second quantization via the construction protocol~\cite{NielsenChuang00}. These calculations show the reduced number of Pauli terms, the total number of quantum circuits in JW is approximately $27$ times that of GC for QuGCM, showing exponential advantage in calculating all values of elements for the process. The details are given in Appendix~\ref{appendix:GATE_comparison}. This compression from JW to GC results in shallower quantum circuits, thereby improving resilience against noise on current hardware.

The ADAPT-GCIM minimizes the number of bases needed by selecting bases dynamically from the gradient value of energy as mentioned in Sec.~\ref{ADAPT-GCIM}. This results in fewer matrix elements to evaluate on the quantum circuit, thereby reducing the number of quantum circuits required. The number of basis reduced in each case due to ADAPT-GCIM for both JW and GC is presented in Table~\ref{tab:comparison_basis_needed}. This clearly shows that fewer basis states spanning our solutions are selected and used. For example, $6$ or $4$ in ADAPT-GCIM, compared to the total of $15$ basis states in JW in the QuGCM approach. In some cases, this even reduces to three from $8$ states for GC encoding in QuGCM. This signifies the advantages of ADAPT-GCIM over QuGCM.

The computed spectra with GC encoding remain close to exact-diagonalization benchmarks, particularly under noisy conditions, highlighting the efficiency of this mapping for variational and GCM-based methods. This advantage arises from the reduced qubit count and lower circuit depth, which mitigate error accumulation. The detailed energy values for this case are summarized alongside results for VQE and ADAPT-VQE under the same scheme in Table \ref{tab:energy_comparison_GC_Ar38}.

\begin{table}[ht]
    \centering
    \caption{Number of generated basis used to obtain energy eigenvalues for different $J^\pi$ states of $^{38}$Ar for QuGCM and ADAPT-GCIM.}
    \begin{tabular}{|c|c|c|c|c|}
        \hline
        & \multicolumn{2}{c|}{\textbf{JW}} & \multicolumn{2}{c|}{\textbf{GC}} \\ \cline{2-5}
        \textbf{$\mathbf{J^\pi}$} & \textbf{QuGCM} & \textbf{ADAPT} & \textbf{QuGCM} & \textbf{ADAPT} \\ 
         &  & \textbf{GCIM} &  & \textbf{GCIM} \\ \hline
        \makebox[8mm][c]{$0_1^+$} & 15 & 6 & 8 & 3 \\ \hline
        $2_1^+$ & 15 & 4 & 4 & 3 \\ \hline
        $2_2^+$ & 15 & 4 & 4 & 3 \\ \hline
        $0_2^+$ & 15 & 4 & 8 & 3 \\ \hline
        $1^+$ & 9 & 4 & 2 & 2 \\ \hline
    \end{tabular}
    \label{tab:comparison_basis_needed}
\end{table}

\begin{table*}[t]
\centering
\caption{Comparison of energy eigenvalues for different $J^\pi$ states of $^{38}$Ar using VQE, ADAPT-VQE~\cite{singh2025advancingquantumsimulationsnuclear}, and calculated using QuGCM for GC, with and without noise. Exact diagonalization values are shown for reference.}
\label{tab:energy_comparison_GC_Ar38}
\begin{tabular}{|c|c|c|c|c|c|c|}
\hline
\textbf{$\mathbf{J^\pi}$} & {\textbf{Exact}} & \textbf{Noise} & {\textbf{VQE}} & {\textbf{ADAPT}} & {\textbf{QuGCM}} & {\textbf{ADAPT}} \\
 & & \textbf{level} & & {\textbf{VQE}} & & {\textbf{GCIM}} \\
\hline
\multirow{2}{*}{$0_1^+$} & -152.677 & Without Noise & -151.7937(4477) & -152.3999(8798) & -152.677(0) & -152.672(5) \\
\cline{3-7}
 & & With Noise & -139.8726(94223) & -141.6124(72305) & -154.146(1469) & -154.146(1469) \\
\hline
\multirow{2}{*}{$2_1^+$} & -151.093 & Without Noise & -151.0934(18) & -150.8583(11249) & -151.093(0) & -151.093(0) \\
\cline{3-7}
 & & With Noise & -149.6657(5218) & -142.1343(45646) & -150.463(630) & -150.885(208) \\
\hline
\multirow{2}{*}{$2_2^+$} & -149.700 & Without Noise & -149.7003(0) & -149.7034(1) & -149.700(0) & -149.700(0) \\
\cline{3-7}
 & & With Noise & -149.7002(2) & -149.5180(5566) & -150.694(994) & -150.887(1187) \\
\hline
\multirow{2}{*}{$0_2^+$} & -149.225 & Without Noise & -137.4674(187408) & -140.6414(144897) & -149.225(0) & -149.225(0) \\
\cline{3-7}
 & & With Noise & -110.8731(305988) & -129.7738(210019) & -149.790(565) & -149.792(567) \\
\hline
\multirow{2}{*}{$1^+$} & -149.104 & Without Noise & -149.8602(6486) & -148.5788(42723) & -149.104(0) & -149.104(0) \\
\cline{3-7}
 & & With Noise & -136.1554(45441) & -141.8163(20340) & -145.313(3791) & -145.551(3553) \\
\hline
\end{tabular}
\end{table*}

\begin{figure}[ht]
    \centering
    \includegraphics[width=\linewidth]{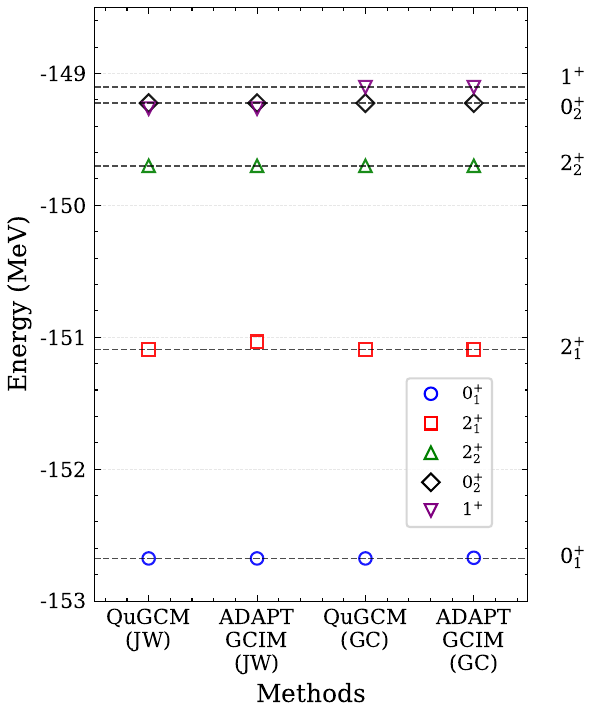}
    \caption{Comparison of energies of levels of $^{38}$Ar for QuGCM and ADAPT-GCIM using JW and GC encodings on noiseless simulator, values are as obtained in Tables \ref{tab:energy_comparison_Ar38_JW} and \ref{tab:energy_comparison_GC_Ar38}.}
    \label{fig:energy_JWvsGC_quGCM_Ar38}
\end{figure}

\begin{figure}[ht]
    \centering
    \includegraphics[width=\linewidth]{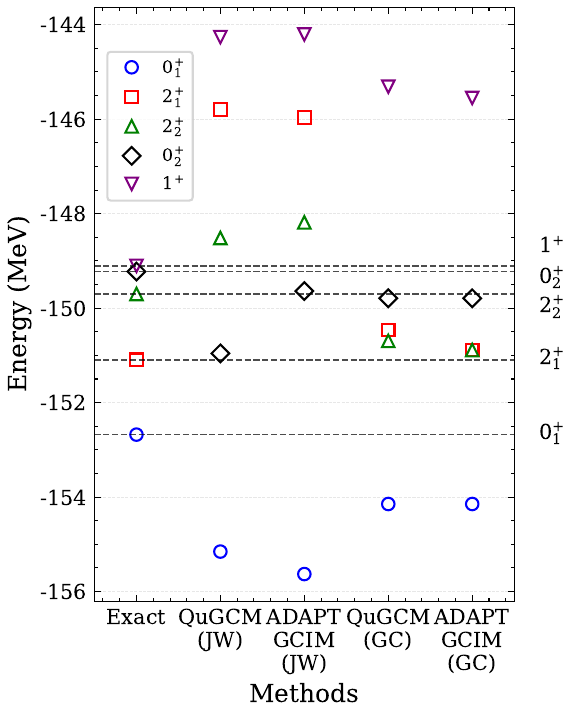}
    \caption{Comparison of energies of levels of $^{38}$Ar for QuGCM and ADAPT-GCIM using JW and GC encodings with noise, values are as obtained in Tables \ref{tab:energy_comparison_Ar38_JW} and \ref{tab:energy_comparison_GC_Ar38}.}
    \label{fig:energy_comp_JWvsGC_quGCM_Ar38_withnoise}
\end{figure}

From Tables \ref{tab:energy_comparison_Ar38_JW} and  \ref{tab:energy_comparison_GC_Ar38}, we see that in the noiseless limit, both QuGCM and ADAPT-GCIM reproduce the exact diagonalization energies. Under the simulated noise model, the QuGCM-based energies shift away from the exact values; the JW results exhibit larger run-to-run scatter (larger reported MAD) for several states, while the GC-encoded results generally remain closer to the exact values. The obtained values for both mappings is compared and illustrated in Fig.~\ref{fig:energy_JWvsGC_quGCM_Ar38} and \ref{fig:energy_comp_JWvsGC_quGCM_Ar38_withnoise} for noiseless and noisy simulators, signifying consistency of ADAPT-GCIM to closely track the QuGCM central values in most cases even for both the encodings, with only modest, state-dependent differences under noise.

The consistency of both encodings is validated by comparing the obtained values for energy spectra of $^{38}$Ar nuclei under both JW and GC schemes. This consistency of GC for same results allows selective access to relevant subspaces without introducing large numbers of Pauli terms. However, operator construction differs from UCC in JW and requires new gate decompositions. Despite this, this remains favorable with GC encoding mirroring the role of UCC operators but with compressed forms that scale better with qubit resources.

\subsection{\texorpdfstring{$^{6}$Li}{6Li}}
It is evident that the JW transformation scales inefficiently with the size of the system. As the problem becomes larger, the number of required qubits and excitation operators increases significantly. This leads to deeper quantum circuits, which are more susceptible to noise and errors, resulting in a substantial increase in computational resource demands. Mitigating this issue is discussed here with example of the GC scheme directly employed for $^6$Li. Within this framework, the basis states are reorganized according to a binary-reflected ordering. This reordering enables a more compact representation of the excitation operators.

\begin{table*}[t]
\centering
\caption{Comparison of energy eigenvalues for different $J^\pi$ states of $^{6}$Li using VQE, ADAPT-VQE~\cite{singh2025advancingquantumsimulationsnuclear}, and values calculated using QuGCM and ADAPT-GCIM for GC encoding, with and without noise. Exact diagonalization values are shown for reference.}
\label{tab:energy_comparison_Li6_final}
\begin{tabular}{|c|c|c|c|c|c|c|}
\hline
\textbf{$\mathbf{J^\pi}$} & {\textbf{Exact}} & \textbf{Noise level} & {\textbf{VQE}} & {\textbf{ADAPT-VQE}} & {\textbf{QuGCM}} & {\textbf{ADAPT-GCIM}} \\
\hline
\multirow{2}{*}{$1^+_1$} & -5.433 & Without Noise & -5.3757(1555) & -5.4254(459) & -5.433(0) & -5.433(0) \\
\cline{3-7}
 & & With Noise & -4.9938(13217) & -5.1101(10181) & -5.419(14) & -5.424(9) \\
\hline
\multirow{2}{*}{$3^+$} & -5.009 & Without Noise & -5.0099(2) & -5.0089(0) & -5.009(0) & -5.009(0) \\
\cline{3-7}
 & & With Noise & -5.0089(0) & -5.0089(1) & -4.876(133) & -4.880(129) \\
\hline
\multirow{2}{*}{$0^+$} & -3.910 & Without Noise & -3.0626(6780) & -3.6782(5792) & -3.332(578) & -3.332(578) \\
\cline{3-7}
 & & With Noise & -2.8558(15756) & -3.1198(13085) & -3.137(773) & -3.964(54) \\
\hline
\multirow{2}{*}{$1^+_2$} & -1.273 & Without Noise & -1.1941(815) & -1.2331(9776) & -1.311(38) & -1.311(38) \\
\cline{3-7}
 & & With Noise & -1.1560(10519) & -1.1814(8276) & -1.146(127) & -1.135(118) \\
\hline
\multirow{2}{*}{$2^+_1$} & -0.510 & Without Noise & -0.5097(2) & -0.5108(18) & -0.5099(0) & -0.510(0) \\
\cline{3-7}
 & & With Noise & -0.5013(524) & -0.4925(789) & -0.5009(519) & -0.515(5) \\
\hline
\multirow{2}{*}{$2^+_2$} & 0.632 & Without Noise & 0.6321(2) & 0.6309(212) & 0.6320(2) & 0.6322(0) \\
\cline{3-7}
 & & With Noise & 0.6211(109) & 0.6021(301) & 0.6264(58) & 0.6303(19) \\
\hline
\end{tabular}
\end{table*}

We treat $^{6}$Li as a system of $^{4}$He core with one valence proton and one valence neutron in the $p$-shell. The Cohen–Kurath nuclear-nuclear interaction is used to generate the Hamiltonian. For the Jordan–Wigner transformation with the $J$-scheme, only $N=4$ qubits are required, whereas the full $m$-scheme demands $N=12$ qubits. The substantially larger qubit requirement reflects the increased dimensionality of the Hilbert space spanned by the system. To resolve this, the problem is reformulated using the GC encoding, which reduces the qubit requirement to $N=4$ even within the m-scheme. The states in qubit form and individual terms of the Hamiltonian formed for the $0^+$ state in the sum of sparse Pauli matrices are given in Appendix~\ref{app:Li6}.
Using this, the $0^+$, $1^+$, $2^+$, and $3^+$ states are studied, and results are presented in the Table~\ref{tab:energy_comparison_Li6_final}. It demonstrates that the results of QuGCM and ADAPT-GCIM are less prone to errors due to noise, owing to their shallower circuits, and show better convergence for the $0^+$ and $1^+$ states. For higher states, the difference between QuGCM and standard VQE becomes negligible.

\begin{table}[!hbt]
    \centering
    \caption{Number of generated basis used to obtain energy eigenvalues for different $J^\pi$ states of $^{6}$Li for QuGCM and ADAPT-GCIM.}
    \begin{tabular}{|c|c|c|c|c|}
        \hline
        \textbf{$J^\pi$} & \textbf{QuGCM} & \textbf{ADAPT} \\ 
         &  & \textbf{GCIM} \\ \hline
        \makebox[8mm][c]{$0_1^+$}  & 8 & 6 \\ \hline
        $3^+$ & 15 & 3 \\ \hline
        $0^+$ & 10 & 6 \\ \hline
        $1_2^+$ & 8 & 3 \\ \hline
        $2_1^+$ & 9 & 3 \\ \hline
        $2_2^+$ & 9 & 2 \\ \hline
    \end{tabular}
    \label{tab:comparison_generatedbasis_for_Lithium}
\end{table}

\begin{figure*}[th]
    \centering
    \includegraphics[width=\linewidth]{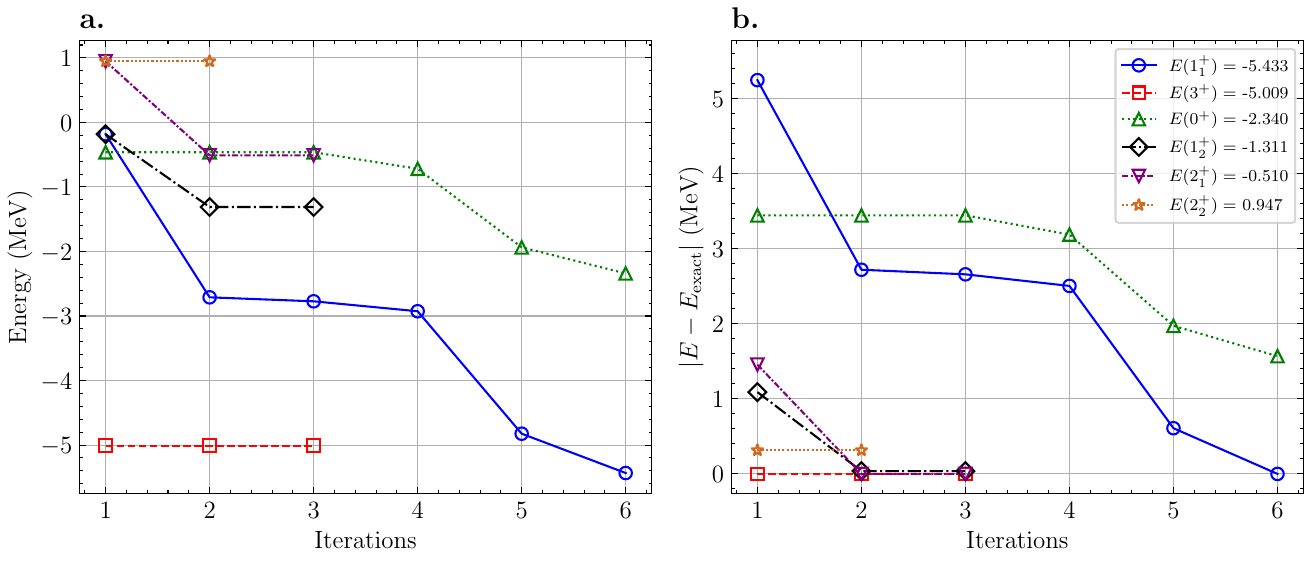}
    \caption{\textbf{a.} Plot showing convergence of energies of different states of $^6$Li nuclei with number of iterations and variation of energy with increase in Hilbert subspace for noiseless simulations. \textbf{b.} Deviation and approach of the energy values to the exact values by plotting the difference.}
    \label{fig:energy_convergence_Li6_iterations}
\end{figure*}

\begin{figure*}[th]
    \centering
    \includegraphics[width=\linewidth]{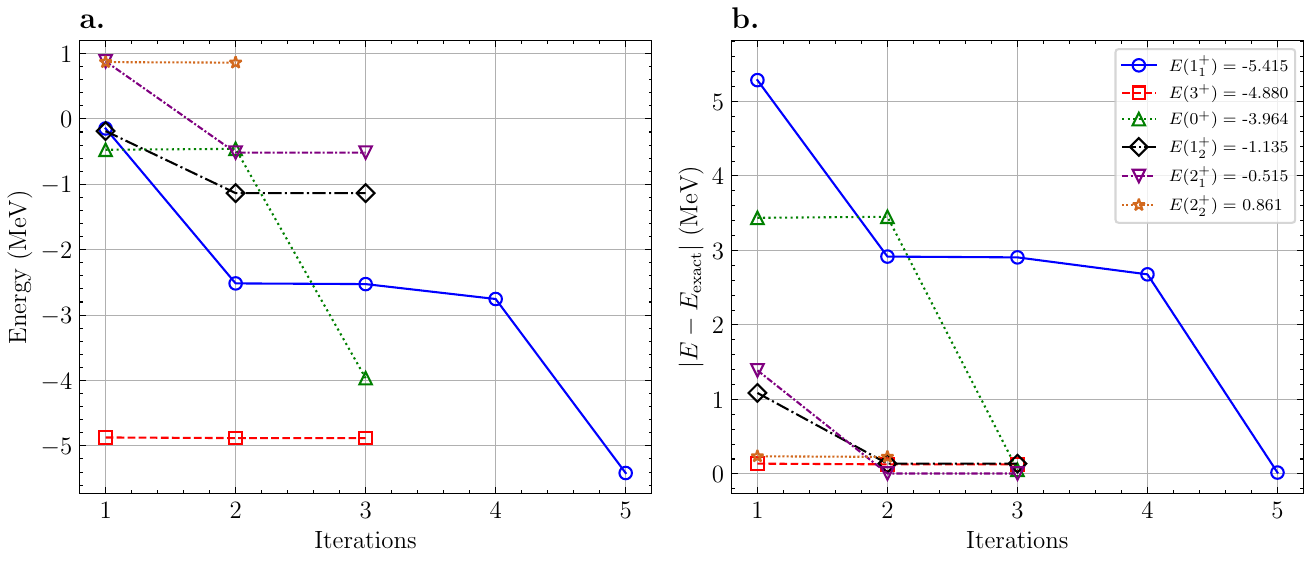}
    \caption{Similar to Fig.~\ref{fig:energy_convergence_Li6_iterations}, plot showing convergence of energies of different states of $^6$Li nuclei with noise.}
    \label{fig:with_noise_energy_convergence_Li6_iterations}
\end{figure*}

Similar to Table~\ref{tab:comparison_basis_needed}, we have shown the number of bases required for ADAPT-GCIM in comparison to QuGCM in Table~\ref{tab:comparison_generatedbasis_for_Lithium}, which shows efficient resource usage when compared to the QuGCM method by reducing the need for generated bases. As observed, there is a significant drop in the required basis to three or two for the $2_1^+$ and $2_2^+$ states, showing the advantage of the ADAPT-GCIM method for accurately finding eigenvalues with smaller subspaces than all possible basis configurations for all states. For $^6$Li, we have plotted the convergence of energy with iteration in Fig.~\ref{fig:energy_convergence_Li6_iterations} and \ref{fig:with_noise_energy_convergence_Li6_iterations} for noiseless and noisy simulations, respectively. From both  results, it is clear that we have achieved the convergence with significant accuracy.

\section{Discussion} \label{Discussion}
The QuGCM provides a non-variational approach for finding the eigenvalues. The linear combination of basis states forming the final eigenstate is determined not by iterative classical optimization as in VQE, but by using the UCC-style operators to systematically build a non-orthogonal generator basis and solving the generalized eigenvalue problem of the Hill-Wheeler-Griffin equation. This makes the QuGCM method optimization free, unlike VQE. This methodology inherently avoids the primary challenges of variational methods, including optimization-induced errors, the barren-plateau phenomenon, and the tendency to converge to local minima. Furthermore, the absence of parameterized gates reduces overall circuit depth and, consequently, the impact of noise. 
When it comes to adaptive methods, ADAPT-VQE improves the VQE ansatz by including additional basis states, thereby constructing the eigenstate. ADAPT-GCIM is also similar to ADAPT-VQE, which includes basis states with larger contributions to energy levels per quantum circuit. However, it does not superimpose all of them into a single circuit as done in ADAPT-VQE. ADAPT-GCIM also retains QuGCM's parameter-free operation and is substantially faster and more robust than its variational counterpart ADAPT-VQE. Because fewer gates are involved in quantum circuits for GCM-based methods, the contribution of noise from each gate decreases, leading to overall noise reduction. This makes QuGCM and ADAPT-GCIM less prone to noise errors than their variational counterparts, VQE and ADAPT-VQE, respectively.
ADAPT-GCIM uses fewer but more significant generator coordinates than QuGCM, with fewer iterations at times. This makes it computationally more accessible by finding eigenvalues in a lower subspace of the complete configuration space. The reduced number of needed basis is reported in Tables~\ref{tab:comparison_basis_needed} and~\ref{tab:comparison_generatedbasis_for_Lithium}. This results in a lower need for quantum circuits in ADAPT-GCIM, making it more resource-efficient and less noisy than QuGCM. This is highly evident from plots of convergences shown in Fig.~\ref{fig:convergence}, and ~\ref{fig:energy_convergence_Li6_iterations}. It therefore gives satisfactory values for noise as seen in Fig.~\ref{fig:with_noise_energy_convergence_Li6_iterations}.  Due to significant exploration of reasonable subspace because of consecutive operators on the state, ADAPT-GCIM is more capable of getting close to more accurate values, as shown for higher energy levels $1^+$ and $0_2^+$ in tables~\ref{tab:energy_comparison_Ar38_JW} and ~\ref{tab:energy_comparison_GC_Ar38}. The results are more accurate and closer to the exact eigenvalues for all energy levels in both noisy and noiseless cases. The reasonable and close energy values are obtained in most of the cases as noted in tables ~\ref{tab:tablefordeuteron},~\ref{tab:energy_comparison_Ar38_JW}, ~\ref{tab:energy_comparison_GC_Ar38}, and~\ref{tab:energy_comparison_Li6_final}.
The efficiency of both QuGCM and ADAPT-GCIM is significantly amplified when paired with GC encoding. In contrast to the JW transformation, GC mapping requires fewer qubits and translates fermionic operators into Pauli expressions, resulting in substantially smaller circuits as calculated in Appendix~\ref{appendix:GATE_comparison}. This directly yields shallower quantum circuits with reduced gate counts, mitigating decoherence and gate errors.  As a result, energy calculations using the combination of ADAPT-GCIM and GC encoding achieve higher accuracy and lower gate counts than the JW-based implementation. The efficiency of GC over JW is observed with more accurate results in Sec.~\ref{Results}.  This is shown by the reduced amount of noise for GC calculations in Fig.~\ref{fig:energy_comp_JWvsGC_quGCM_Ar38_withnoise} and significantly closed values obtained for both noise and noiseless case in Table~\ref{tab:energy_comparison_Li6_final}. Therefore, combining efficient non-variational GCM-based algorithms with a resource-optimized GC encoding offers a promising pathway toward practical quantum simulations on near-term quantum devices.

\section{Conclusion}
\label{Conclusion}
In this work, we have presented the performance of the adaptive, parameter-free QuGCM and ADAPT-GCIM. Our analysis shows the advantage of this non-variational approach, which circumvents the inherent challenges of variational algorithms, such as barren plateaus and convergence to local minima. By eliminating the need for a classical optimization loop, these GCM-based methods provide a more stable and direct path to the target eigenstate, representing a significant improvement in reliability and computational efficiency.

Furthermore, we have explored a highly efficient and robust methodology for performing nuclear shell model calculations on a quantum computer by significantly reducing resource requirements. We have demonstrated that mapping many-body states using GC encoding outperforms the conventional JW transformation. This approach yields a considerable reduction in the number of qubits and the circuit depth required for simulation. The resulting efficiency not only enhances the accuracy of results by reducing noise but also makes quantum calculations more practical for complex nuclear systems. The synergy between a resource-sparing encoding and an advanced algorithm is a key finding of this study.

One major contribution of this work is the formal establishment of the QuGCM as a quantum analog of GCM, bridging a critical gap between established nuclear theory and quantum computational techniques. The classical GCM, a powerful beyond-mean-field technique, is computationally limited by the size of the basis it can handle. However, its quantum analogue, QuGCM, may overcome this barrier by leveraging quantum hardware to manage vastly larger configuration spaces. This can be used to study significant correlations and collective phenomena in nuclei that are classically intractable, marking a significant step forward for high-precision nuclear structure calculations.

Overall, this work highlights the potential of resource-efficient GC encoding coupled with non-variational GCM-based algorithms to scale quantum simulations of nuclear systems. The promising results presented here pave the way for several future lines of inquiry. An immediate next step is to apply this methodology as an alternative to classical GCM in nuclear or many-body physics, such as to determine collective behavior from the many-body system of individual particles. Benchmarking the performance and inherent error resilience of the ADAPT-GCIM and GC frameworks on real quantum hardware is also an essential direction for validating its practical utility. Further theoretical work could involve extending the framework to compute additional nuclear observables, such as electromagnetic transition rates, and developing more sophisticated generator selection criteria to accelerate convergence for other methods as well. 

\section{Acknowledgment}
This work is supported by the SERB-DST, Govt.~of India, via project \sloppy{CRG/2022/009359}.
We acknowledge the National Supercomputing Mission (NSM) for providing computing resources of `PARAM Ganga' at IIT Roorkee, which is implemented by C-DAC and supported by MeitY and DST, Govt.~of India.

\begin{appendix}
\section{Parameter independence of results using QuGCM.}
\label{appendix_parameter_indpendence}
\begin{figure}[th]
    \centering
    \includegraphics[width=\linewidth]{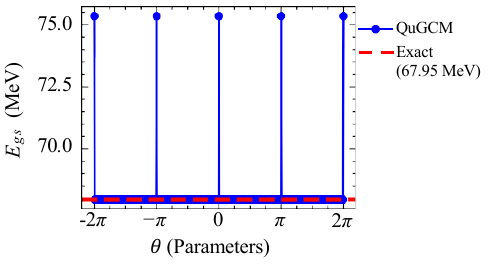}
    \caption{Variation of the deuteron ground-state energy with respect to $\theta$ in the Generation operators. The energy remains invariant under different choices of $\theta$, confirming the robustness of the method. The abrupt values at multiples of $\pi$ are due to the formation of non-singular norm matrix $\mathcal{N}$ in Hill-Wheeler equation $\mathcal{H} \psi = E\mathcal{N}\psi$.}
    \label{fig:theta_variation}
\end{figure}

For the deuteron, generator bases are prepared by superposition of different bases using a single parameter $\theta$ for the QuGCM. The energy converges to the same value irrespective of $\theta$, confirming that the ansatz is insensitive to arbitrary parameter initialization. This invariance underscores the stability of the QuGCM and ADAPT-GCIM approaches and ensures reliable convergence without the need to fine-tune initial parameter values. As seen in Fig.~\ref{fig:theta_variation}, we calculated the $E_{gs}$ for different values of $\theta$ ranging from $-2\pi$ to $2\pi$, which confirms the fact mentioned above. The abrupt values at multiples of $\pi$ are due to the formation of a non-singular norm matrix $\mathcal{N}$ in Hill-Wheeler Eq.~\ref{HillWheelerEquation} that makes it unsolvable.
\section{\texorpdfstring{$^{38}$Ar States and Qubit Mapping}{38Ar States and Qubit Mapping}}\label{app:Ar38}

The calculations for the $^{38}$Ar isotope are performed within the $sd$ valence shell model space, where the relevant physics is described by the interactions of nucleons in the $1s_{1/2}$, $0d_{3/2}$, and $0d_{5/2}$ orbitals~\cite{singh2025advancingquantumsimulationsnuclear}. The Qubit mapping from the JW transformation of OH encoding to GC encoding is done to decrease the number of qubits needed. Table~\ref{tab:mapping_Ar38_JW_GC} illustrates the mapping from the 6-qubit Jordan–Wigner (JW) representation of the $0^+$ ground state of $^{38}$Ar to the 3-qubit GC encoding. It further provides a one-to-one correspondence between the basis states in the two encodings.
\begin{table}[ht]
\centering
\caption{Mapping from Jordan-Wigner (JW) basis states to GC qubit states for $^{38}$Ar ($0^+$).}
\label{tab:mapping_Ar38_JW_GC}
\setlength{\tabcolsep}{12pt}
\renewcommand{\arraystretch}{1.2}
\begin{tabular}{@{}ccc@{}}
\toprule
\textbf{State} & \textbf{JW State} & \textbf{GC State} \\
\textbf{Index} & \textbf{(6 Qubits)} & \textbf{(3 Qubits)} \\
\midrule
1 & $\ket{010010}$ & $\ket{000}$ \\
2 & $\ket{001010}$ & $\ket{001}$ \\
3 & $\ket{010001}$ & $\ket{011}$ \\
4 & $\ket{001001}$ & $\ket{010}$ \\
5 & $\ket{100100}$ & $\ket{110}$ \\
\bottomrule
\end{tabular}
\end{table}

Using these bases, the Hermitian operator is constructed using the Single Particle Energies and Two Body Matrix Elements. It is then further reduced into Sparse Pauli form for execution on quantum devices. This explicit Pauli representation of the Hamiltonian for the $0^{+}$ state of $^{38}$Ar, along with the corresponding coefficients obtained using the JW transformation and the GC encoding, are presented in Tables~\ref{tab:H0+arjw} and \ref{tab:H0+argc}, respectively. A comparison between the two indicates that the GC encoding uses fewer qubits and results in fewer terms.

\begin{table}
    \centering
    \caption{Hamiltonian terms for $^{38}$Ar ($0^+$) in the Jordan-Wigner (JW) scheme.}
    \label{tab:H0+arjw}
    \setlength{\tabcolsep}{4pt}
    \footnotesize
    \begin{tabular}{@{}lr | lr@{}}
        \toprule
        \textbf{Pauli Term} & \textbf{Coeff.} & \textbf{Pauli Term} & \textbf{Coeff.} \\
        \midrule
        $I$ & $-187.949844$ & $ X_1 X_2 Y_4 Y_5$ & $-0.071698$ \\
        $ X_0 X_1 X_3 X_4$ & $0.112616$ & $ X_1 Y_2 X_3 Z_4 Y_5$ & $-0.017452$ \\
        $ X_0 Y_1 X_3 Y_4$ & $-0.030779$ & $ X_1 Y_2 X_4 Y_5$ & $0.071698$ \\
        $ X_0 Z_1 X_2 X_3 Z_4 X_5$ & $0.089998$ & $ Y_1 X_2 X_3 Z_4 Y_5$ & $0.017452$ \\
        $ X_0 Z_1 X_2 X_4 X_5$ & $-0.017452$ & $ Y_0 Z_1 X_2 Y_3 Z_4 X_5$ & $-0.089998$ \\
        $ X_0 Z_1 Y_2 X_3 Z_4 Y_5$ & $-0.089998$ & $ Y_1 X_2 X_4 Y_5$ & $-0.071698$ \\
        $ X_0 Z_1 Y_2 X_4 Y_5$ & $0.017452$ & $ Z_1$ & $74.714645$ \\
        $ Y_0 X_1 X_3 Y_4$ & $0.030779$ & $ Z_1 Z_3$ & $-37.357837$ \\
        $ Y_0 Y_1$ & $0.069808$ & $ Z_1 Z_4$ & $-37.356807$ \\
        $ Y_0 Y_1 Y_3 Y_4$ & $0.112616$ & $ Z_2$ & $37.938686$ \\
        $ Y_0 Z_1 X_2 Y_4 X_5$ & $0.017452$ & $ Z_2 Z_5$ & $-37.938686$ \\
        $ Y_0 Z_1 Y_2 Y_3 Z_4 Y_5$ & $0.089998$ & $ X_3 X_4$ & $-0.069808$ \\
        $ X_1 X_2 X_3 Z_4 X_5$ & $0.017452$ & $ Y_3 Y_4$ & $-0.069808$ \\
        $ X_1 X_2 X_4 X_5$ & $-0.071698$ & $ Z_3$ & $75.296516$ \\
        $ X_0 X_1$ & $0.069808$ & $ Z_4$ & $74.714641$ \\
        $ X_0 X_1 Y_3 Y_4$ & $0.112616$ & $ Z_5$ & $37.938686$ \\
        $ X_0 Y_1 Y_3 X_4$ & $0.030779$ & $ Z_0$ & $75.296512$ \\
        $ X_0 Z_1 X_2 Y_3 Z_4 Y_5$ & $0.089998$ & $ Z_0 Z_3$ & $-37.938678$ \\
        $ X_0 Z_1 X_2 Y_4 Y_5$ & $-0.017452$ & $ Z_0 Z_4$ & $-37.357833$ \\
        $ X_0 Z_1 Y_2 Y_3 Z_4 X_5$ & $0.089998$ & $ Y_1 Y_2 X_3 Z_4 X_5$ & $0.017452$ \\
        $ X_0 Z_1 Y_2 Y_4 X_5$ & $-0.017452$ & $ Y_1 Y_2 Y_3 Z_4 Y_5$ & $0.017452$ \\
        $ Y_0 X_1 Y_3 X_4$ & $-0.030779$ & $ Y_1 Y_2 X_4 X_5$ & $-0.071698$ \\
        $ Y_0 Y_1 X_3 X_4$ & $0.112616$ & $ Y_1 Y_2 Y_4 Y_5$ & $-0.071698$ \\
        $ Y_0 Z_1 X_2 X_3 Z_4 Y_5$ & $0.089998$ & $ Y_0 Z_1 Y_2 Y_4 Y_5$ & $-0.017452$ \\
        $ Y_0 Z_1 Y_2 X_3 Z_4 X_5$ & $0.089998$ & $ Y_1 X_2 Y_4 X_5$ & $0.071698$ \\
        $ Y_0 Z_1 Y_2 X_4 X_5$ & $-0.017452$ & $ X_1 X_2 Y_3 Z_4 Y_5$ & $0.017452$ \\
        $ X_1 Y_2 Y_3 Z_4 X_5$ & $0.017452$ & $ Y_1 X_2 Y_3 Z_4 X_5$ & $-0.017452$ \\
        $ X_1 Y_2 Y_4 X_5$ & $-0.071698$ & $ Y_0 Z_1 X_2 X_4 Y_5$ & $-0.017452$ \\
        \bottomrule
    \end{tabular}
\end{table}

\begin{table}
    \centering
    \caption{Hamiltonian terms for $^{38}$Ar ($0^+$) in the GC scheme.}
    \label{tab:H0+argc}
    \setlength{\tabcolsep}{6pt}
    \renewcommand{\arraystretch}{1.2}
    \begin{tabular}{@{}lr | lr@{}}
        \toprule
        \textbf{Pauli Term} & \textbf{Coefficient} & \textbf{Pauli Term} & \textbf{Coefficient} \\
        \midrule
        $I$  & $-93.974922$ & $X_0$ & $-0.034904$ \\
        $ Z_2$ & $-56.036235$ & $X_0 X_2$ & $-0.034904$ \\
        $ Z_1$ & $19.260276$ & $X_2 Z_0$ & $0.179996$ \\
        $ Z_0 Z_2$ & $18.679435$ & $X_1 Z_2$ & $0.225232$ \\
        $ Y_1 Y_2$ & $0.143396$ & $X_0 Z_1$ & $0.034904$ \\
        $ Y_0 Y_1$ & $0.034904$ & $X_0 Z_1 Z_2$ & $0.034904$ \\
        $ Y_1 Y_2 Z_0$ & $0.143396$ & $X_1 X_2 Z_0$ & $-0.143396$ \\
        $ X_2$ & $0.179996$ & $X_1 Y_0 Y_2$ & $0.034904$ \\
        $ X_1$ & $0.225232$ & $X_1 Z_0 Z_2$ & $0.061559$ \\
        $ X_1 X_2$ & $-0.143396$ & $X_2 Y_0 Y_1$ & $0.034904$ \\
        $ X_2 Z_1$ & $-0.179996$ & $X_2 Z_0 Z_1$ & $-0.179996$ \\
        $ Z_1 Z_2$ & $-18.678409$ & $Y_0 Y_1 Z_2$ & $0.034904$ \\
        $ Z_0$ & $-19.259250$ & $Y_0 Y_2 Z_1$ & $0.034904$ \\
        $ X_0 Z_2$ & $-0.034904$ & $X_0 X_1 X_2$ & $0.034904$ \\
        $ Y_0 Y_2$ & $-0.034904$ & $X_0 X_1 Z_2$ & $0.034904$ \\
        $ X_0 X_1$ & $0.034904$ & $X_0 X_2 Z_1$ & $0.034904$ \\
        $ X_1 Z_0$ & $0.061559$ & $X_0 Y_1 Y_2$ & $-0.034904$ \\
        $ Z_0 Z_1$ & $19.260280$ & $Z_0 Z_1 Z_2$ & $-18.678405$ \\
        \bottomrule
    \end{tabular}
\end{table}

\section{\texorpdfstring{$^6$Li States and Qubit Mapping}{6Li}}\label{app:Li6}

\subsection{Valence shell model space}
The $^6$Li nucleus is modeled using the $p$-shell valence space, mapping the Slater determinants of the active space onto qubit registers. The 12-qubit JW states are compressed into 4-qubit GC states as defined in Table~\ref{tab:mapping_Li6_direct}.

\begin{table}
\centering
\caption{Basis mapping for the $^6$Li ($0^+$) state, showing the correspondence between physical states and the 4-qubit GC \textbf{register}.}
\label{tab:mapping_Li6_direct}
\setlength{\tabcolsep}{10pt}
\renewcommand{\arraystretch}{1.1}
\begin{tabular}{@{}clc@{}}
\toprule
\textbf{State} & \textbf{Slater} & \textbf{GC Qubit} \\
\textbf{Index} & \textbf{Determinant} & \textbf{State} \\
$i$ & $|d_i\rangle$ & $|q_3 q_2 q_1 q_0\rangle$ \\
\midrule
1  & $\ket{100000000100}$ & $\ket{0000}$ \\
2  & $\ket{010000000010}$ & $\ket{0001}$ \\
3  & $\ket{001000000010}$ & $\ket{0011}$ \\
4  & $\ket{010000000001}$ & $\ket{0010}$ \\
5  & $\ket{001000000001}$ & $\ket{0110}$ \\
6  & $\ket{000010010000}$ & $\ket{0100}$ \\
7  & $\ket{000001010000}$ & $\ket{1100}$ \\
8  & $\ket{000010001000}$ & $\ket{1000}$ \\
9  & $\ket{000001001000}$ & $\ket{1001}$ \\
10 & $\ket{000100100000}$ & $\ket{1101}$ \\
\bottomrule
\end{tabular}
\end{table}
\pagebreak
\subsection{\texorpdfstring{Hamiltonian for $0^+$ state in GC}{Hamiltonian for 0+ state in GC}}
The explicit Hamiltonian terms for the $^6$Li $0^+$ state under GC encoding are listed in Table\ref{tab:H0+ligc}.
\setlength{\LTcapwidth}{0.95\columnwidth}
\begin{longtable}{lr | lr}
    \caption{Hamiltonian coefficients and Pauli terms for $^{6}$Li ($0^+$) in the GC scheme.} \label{tab:H0+ligc} \\
    
    \toprule
    \textbf{Pauli Term} & \textbf{Coeff.} & \textbf{Pauli Term} & \textbf{Coeff.} \\
    \midrule
    \endfirsthead
    
    \caption[]{Hamiltonian coefficients for $^{6}$Li ($0^+$) (continued)} \\
    \toprule
    \textbf{Pauli Term} & \textbf{Coeff.} & \textbf{Pauli Term} & \textbf{Coeff.} \\
    \midrule
    \endhead
    
    \midrule
    \multicolumn{4}{r}{\textit{Continued ...}} \\
    \endfoot
    
    \bottomrule
    \endlastfoot

    $ I $ & $0.242340$ & $ X_3Z_1 $ & $0.026711$ \\
    $ Z_3 $ & $-0.051668$ & $ Z_1Z_3 $ & $-0.351740$ \\
    $ Z_2 $ & $-0.351740$ & $ X_1Z_2 $ & $0.331685$ \\
    $ X_2Z_3 $ & $-0.135605$ & $ Y_1Y_2 $ & $0.486446$ \\
    $ Y_2Y_3 $ & $-0.149642$ & $ X_1X_2X_3 $ & $-0.135605$ \\
    $ X_1 $ & $0.026711$ & $ X_1X_3Z_2 $ & $0.248567$ \\
    $ X_1X_3 $ & $-0.248567$ & $ X_1Z_2Z_3 $ & $0.331685$ \\
    $ X_3 $ & $0.026711$ & $ X_2Y_1Y_3 $ & $-0.135605$ \\
    $ X_2 $ & $-0.135605$ & $ X_3Y_1Y_2 $ & $-0.135605$ \\
    $ X_2X_3 $ & $0.149642$ & $ Y_1Y_2Z_3 $ & $0.486446$ \\
    $ X_3Z_2 $ & $-0.026711$ & $ Y_2Y_3Z_1 $ & $-0.149642$ \\
    $ Z_2Z_3 $ & $-0.057731$ & $ X_0 $ & $-0.067722$ \\
    $ Z_1 $ & $-0.057731$ & $ X_0X_3 $ & $-0.026711$ \\
    $ X_1Z_3 $ & $0.026711$ & $ X_3Z_0 $ & $-0.133192$ \\
    $ Y_1Y_3 $ & $-0.248567$ & $ Z_0Z_3 $ & $0.191529$ \\
    $ X_1X_2 $ & $-0.408183$ & $ X_0Z_2 $ & $0.411575$ \\
    $ X_2Z_1 $ & $-0.222791$ & $ Y_0Y_2 $ & $-0.261380$ \\
    $ Z_1Z_2 $ & $-0.446855$ & $ X_0X_2X_3 $ & $-0.395699$ \\
    $ X_1X_2Z_3 $ & $-0.408183$ & $ X_0X_3Z_2 $ & $0.026711$ \\
    $ X_1Y_2Y_3 $ & $0.135605$ & $ X_0Z_2Z_3 $ & $0.174856$ \\
    $ X_2X_3Z_1 $ & $0.149642$ & $ X_2Y_0Y_3 $ & $-0.120848$ \\
    $ X_2Z_1Z_3 $ & $-0.222791$ & $ X_3Y_0Y_2 $ & $-0.120848$ \\
    $ X_3Z_1Z_2 $ & $-0.026711$ & $ Y_0Y_2Z_3 $ & $-0.261380$ \\
    $ Y_1Y_3Z_2 $ & $0.248567$ & $ Y_2Y_3Z_0 $ & $-0.506821$ \\
    $ Z_1Z_2Z_3 $ & $-0.152846$ & $ X_0Z_1 $ & $0.174856$ \\
    $ Z_0 $ & $0.601002$ & $ Y_0Y_1 $ & $-0.065300$ \\
    $ X_0Z_3 $ & $0.168995$ & $ X_0X_1X_3 $ & $0.516548$ \\
    $ Y_0Y_3 $ & $0.133192$ & $ X_0X_3Z_1 $ & $-0.026711$ \\
    $ X_0X_2 $ & $0.135605$ & $ X_0Z_1Z_3 $ & $0.411575$ \\
    $ X_2Z_0 $ & $-0.135605$ & $ X_1Y_0Y_3 $ & $-0.139914$ \\
    $ Z_0Z_2 $ & $-0.409472$ & $ X_3Y_0Y_1 $ & $0.139914$ \\
    $ X_0X_2Z_3 $ & $0.135605$ & $ Y_0Y_1Z_3 $ & $-0.065300$ \\
    $ X_0Y_2Y_3 $ & $0.395699$ & $ Y_1Y_3Z_0 $ & $-0.318451$ \\
    $ X_2X_3Z_0 $ & $0.506821$ & $ X_0X_1X_2 $ & $-0.628441$ \\
    $ X_2Z_0Z_3 $ & $-0.135605$ & $ X_0X_2Z_1 $ & $0.222791$ \\
    $ X_3Z_0Z_2 $ & $0.133192$ & $ X_0Z_1Z_2 $ & $-0.378941$ \\
    $ Y_0Y_3Z_2 $ & $-0.133192$ & $ X_1Y_0Y_2 $ & $0.366906$ \\
    $ X_0X_1 $ & $-0.026711$ & $ X_2Y_0Y_1 $ & $0.366906$ \\
    $ X_1Z_0 $ & $0.026711$ & $ Y_0Y_3Z_1 $ & $0.133192$ \\
    $ X_3Y_1Y_2Z_0 $ & $0.242086$ & $ Z_0Z_1Z_3 $ & $-0.409472$ \\
    $ X_0X_1Z_3 $ & $-0.026711$ & $ X_0X_1Z_2 $ & $-0.331685$ \\
    $ X_0Y_1Y_3 $ & $0.516548$ & $ X_0Y_1Y_2 $ & $-0.647896$ \\
    $ X_1X_3Z_0 $ & $-0.318451$ & $ X_1X_2Z_0 $ & $0.210596$ \\
    $ X_1Z_0Z_3 $ & $0.026711$ & $ X_1Z_0Z_2 $ & $-0.065300$ \\
    $ X_3Z_0Z_1 $ & $-0.133192$ & $ X_2Z_0Z_1 $ & $-0.261380$ \\
    $ Y_2Y_3Z_0Z_1 $ & $-0.506821$ & $ Y_0Y_1Y_2Y_3 $ & $-0.242086$ \\
    $ Y_1Y_2Z_0Z_3 $ & $-0.210596$ & $ X_0Z_1Z_2Z_3 $ & $-0.615660$ \\
    $ Y_1Y_3Z_0Z_2 $ & $0.318451$ & $ X_1X_2Z_0Z_3 $ & $0.210596$ \\
    $ Y_0Y_2Z_1Z_3 $ & $-0.135605$ & $ X_1X_2X_3Z_0 $ & $0.242086$ \\
    $ X_3Z_0Z_1Z_2 $ & $0.133192$ & $ X_1X_2Y_0Y_3 $ & $-0.242086$ \\
    $ Y_0Y_3Z_1Z_2 $ & $-0.133192$ & $ X_1X_3Y_0Y_2 $ & $-0.242086$ \\
    $ Y_0Y_1Z_2Z_3 $ & $0.026711$ & $ X_1X_3Z_0Z_2 $ & $0.318451$ \\
    $ X_1Y_0Y_2Z_3 $ & $0.366906$ & $ X_0X_2X_3Z_1 $ & $-0.395699$ \\
    $ X_1Y_0Y_3Z_2 $ & $0.139914$ & $ X_0X_2Y_1Y_3 $ & $0.135605$ \\
    $ X_1Y_2Y_3Z_0 $ & $-0.242086$ & $ X_0X_2Z_1Z_3 $ & $0.222791$ \\
    $ X_1Z_0Z_2Z_3 $ & $-0.065300$ & $ X_0X_3Y_1Y_2 $ & $0.135605$ \\
    $ X_2X_3Y_0Y_1 $ & $0.242086$ & $ X_0X_3Z_1Z_2 $ & $0.026711$ \\
    $ X_2Y_0Y_1Z_3 $ & $0.366906$ & $ X_0Y_1Y_2Z_3 $ & $-0.647896$ \\
    $ X_2Y_0Y_3Z_1 $ & $-0.120848$ & $ X_0Y_1Y_3Z_2 $ & $-0.516548$ \\
    $ X_2X_3Z_0Z_1 $ & $0.506821$ & $ X_0Y_2Y_3Z_1 $ & $0.395699$ \\
    $ X_2Y_1Y_3Z_0 $ & $0.242086$ & $ Z_0Z_1Z_2Z_3 $ & $0.601002$ \\
    $ X_2Z_0Z_1Z_3 $ & $-0.261380$ & $ X_0X_1X_2X_3 $ & $0.135605$ \\
    $ X_3Y_0Y_1Z_2 $ & $-0.139914$ & $ X_0X_1X_2Z_3 $ & $-0.628441$ \\
    $ X_3Y_0Y_2Z_1 $ & $-0.120848$ & $ Y_0Y_1Z_2 $ & $0.026711$ \\
    $ X_0X_1X_3Z_2 $ & $-0.516548$ & $ Z_0Z_1Z_2 $ & $0.191529$ \\
    $ X_0X_1Y_2Y_3 $ & $-0.135605$ & $ Y_1Y_2Z_0 $ & $-0.210596$ \\
    $ X_0X_1Z_2Z_3 $ & $-0.331685$ & $ Y_0Y_2Z_1 $ & $-0.135605$ \\
\end{longtable}
\section{Comparison of Quantum gates in JW and GC encoding}
\label{appendix:GATE_comparison}
Implementation costs for fermionic operators are evaluated under JW and GC mappings, demonstrating significant resource savings with the GC scheme. To calculate the element $H_{ij}$ and $N_{ij}$, an operator composed of $G_j^\dagger H G_i$ and $G_j^\dagger G_i$ is used, respectively. These operators are formed by the composition of each operator $G_i$ or $H$, which is already in SparsePauli form, and hence are themselves in that form. $G_i$, although developed using Trotterization as a quantum circuit, is later treated as an operator and converted to sparsePauli form during computation, and therefore has only single Pauli terms and no CNOT gates. These are then measured using ground state or initial basis $\ket{\phi_{HF}}$, where $X$ gates denote the presence of a particle as digit $1$ in the Slater determinant; e.g., $\ket{010}$ implies the ansatz should be just an $X$ gate on the middle qubit. So, the efficiency and gate counts depend on the number of Pauli matrices in each sparse Pauli term. Therefore, the process of finding elements on a quantum computer completely excludes the use of CNOT gates for the basic execution mechanism used in modern quantum hardware. Finally, the number of terms in the sparse Pauli form of each element indicates how many quantum circuits are needed in the process.
\begin{align*}
    \text{Quantum Circuits} = &\ \text{No. of Pauli terms}\\
    \text{No. of Quantum Gates} = & \text{ No. of Pauli terms} \cross
    \\ & \text{total products in each term}
\end{align*}
Using the above formulation, the number of quantum circuits for GC is approximately $27$ times that for JW.
However, these are only single-qubit gates, which don't contribute much to the circuit depth compared to VQE or ADAPT-VQE, where the ansatz itself has highly coherent, depth-increasing CNOT gates. When the GC is compared to JW in this case, it turns out faster and more efficient with a reduced number of quantum gates and sparsePauli terms in operator implication during measurement on the quantum circuit.
For comparison of QuGCM, we present the tables~\ref{tab:gatecountsforGC} and ~\ref{tab:gatecountsforJW}, that tells how many terms are present during the calculation of each element for the simplest case of $^{38}\text{Ar}$ ($0^+$) state, which is a specifically reproduced case from both JW and GC to do a fair comparison. For JW and GC encoding, this Hamiltonian has $56$ terms and $34$ terms, respectively. The number of terms per element ranges from $0$ to $32$ for GC, which is far fewer than $0$ to $352$ for JW. In total, $2126$ circuits are used for GC, which is too few compared to $56694$ in JW. In VQE and ADAPT-VQE, the ansatz consists of many CNOT gates, leading to high circuit depth and high susceptibility to hardware noise. This is completely avoided in the GCM-based approach, as no CNOT gates are used.

\begin{table}[t]
\centering
\[
P_H^{(\mathrm{GC})} =
\begin{pmatrix}
18 & 26 & 29 & 26 & 28 & 28 & 29 & 26 \\
26 & 18 & 32 & 32 & 32 & 32 & 32 & 32 \\
29 & 32 & 18 & 32 & 32 & 32 & 32 & 32 \\
26 & 32 & 32 & 18 & 32 & 32 & 32 & 32 \\
28 & 32 & 32 & 32 & 18 & 32 & 32 & 32 \\
28 & 32 & 32 & 32 & 32 & 18 & 32 & 32 \\
29 & 32 & 32 & 32 & 32 & 32 & 18 & 32 \\
26 & 32 & 32 & 32 & 32 & 32 & 32 & 18
\end{pmatrix}
\]

\[
P_N^{(\mathrm{GC})} =
\begin{pmatrix}
0 & 1 & 2 & 1 & 2 & 4 & 2 & 1 \\
1 & 0 & 4 & 2 & 4 & 8 & 4 & 2 \\
2 & 4 & 0 & 4 & 8 & 8 & 8 & 4 \\
1 & 2 & 4 & 0 & 4 & 8 & 4 & 2 \\
2 & 4 & 8 & 4 & 0 & 8 & 8 & 4 \\
4 & 8 & 8 & 8 & 8 & 0 & 8 & 8 \\
2 & 4 & 8 & 4 & 8 & 8 & 0 & 4 \\
1 & 2 & 4 & 2 & 4 & 8 & 4 & 0
\end{pmatrix}
\]
\caption{
Number of Pauli terms contributing to the Hamiltonian matrix
elements $H_{ij}$ and overlap matrix elements $N_{ij}$ for the
$^{38}\mathrm{Ar}(0^+)$ state using the GC mapping.
}
\label{tab:gatecountsforGC}
\end{table}

\begin{table*}[htbp!]
\centering
\[
P_N^{(\mathrm{JW})} =
\left(
\begin{array}{cccccccccccccccc}
56  & 94  & 88  & 62  & 94  & 94  & 62  & 88  & 94  & 144 & 144 & 152 & 128 & 144 & 144 \\
94  & 93  & 272 & 176 & 240 & 280 & 208 & 280 & 232 & 224 & 290 & 250 & 344 & 352 & 384 \\
88  & 272 & 88  & 176 & 272 & 280 & 176 & 154 & 280 & 260 & 338 & 338 & 200 & 338 & 260 \\
62  & 176 & 176 & 60  & 176 & 208 & 120 & 176 & 208 & 272 & 224 & 256 & 160 & 224 & 272 \\
94  & 240 & 272 & 176 & 93  & 232 & 208 & 280 & 280 & 384 & 352 & 250 & 344 & 290 & 224 \\
94  & 280 & 280 & 208 & 232 & 93  & 176 & 272 & 240 & 290 & 224 & 250 & 344 & 384 & 352 \\
62  & 208 & 176 & 120 & 208 & 176 & 60  & 176 & 176 & 224 & 272 & 256 & 160 & 272 & 224 \\
88  & 280 & 154 & 176 & 280 & 272 & 176 & 88  & 272 & 338 & 260 & 338 & 200 & 260 & 338 \\
94  & 232 & 280 & 208 & 280 & 240 & 176 & 272 & 93  & 352 & 384 & 250 & 344 & 224 & 290 \\
144 & 224 & 260 & 272 & 384 & 290 & 224 & 338 & 352 & 157 & 320 & 272 & 269 & 268 & 299 \\
144 & 290 & 338 & 224 & 352 & 224 & 272 & 260 & 384 & 320 & 157 & 272 & 269 & 299 & 268 \\
152 & 250 & 338 & 256 & 250 & 250 & 256 & 338 & 250 & 272 & 272 & 134 & 172 & 272 & 272 \\
128 & 344 & 200 & 160 & 344 & 344 & 160 & 200 & 344 & 269 & 269 & 172 & 120 & 269 & 269 \\
144 & 352 & 338 & 224 & 290 & 384 & 272 & 260 & 224 & 268 & 299 & 272 & 269 & 157 & 320 \\
144 & 384 & 260 & 272 & 224 & 352 & 224 & 338 & 290 & 299 & 268 & 272 & 269 & 320 & 157 
\end{array}
\right)
\]
\[
P_N^{(\mathrm{JW})} =
\left(
\begin{array}{cccccccccccccccc}
0 & 2 & 2 & 2 & 2 & 2 & 2 & 2 & 2 & 8 & 8 & 8 & 8 & 8 & 8 \\
2 & 0 & 8 & 8 & 8 & 8 & 8 & 8 & 8 & 16 & 16 & 16 & 32 & 32 & 32 \\
2 & 8 & 0 & 8 & 8 & 8 & 8 & 8 & 8 & 16 & 32 & 32 & 16 & 32 & 16 \\
2 & 8 & 8 & 0 & 8 & 8 & 8 & 8 & 8 & 32 & 16 & 32 & 16 & 16 & 32 \\
2 & 8 & 8 & 8 & 0 & 8 & 8 & 8 & 8 & 32 & 32 & 16 & 32 & 16 & 16 \\
2 & 8 & 8 & 8 & 8 & 0 & 8 & 8 & 8 & 16 & 16 & 16 & 32 & 32 & 32 \\
2 & 8 & 8 & 8 & 8 & 8 & 0 & 8 & 8 & 16 & 32 & 32 & 16 & 32 & 16 \\
2 & 8 & 8 & 8 & 8 & 8 & 8 & 0 & 8 & 32 & 16 & 32 & 16 & 16 & 32 \\
2 & 8 & 8 & 8 & 8 & 8 & 8 & 8 & 0 & 32 & 32 & 16 & 32 & 16 & 16 \\
8 & 16 & 16 & 32 & 32 & 16 & 16 & 32 & 32 & 0 & 32 & 32 & 32 & 64 & 32 \\
8 & 16 & 32 & 16 & 32 & 16 & 32 & 16 & 32 & 32 & 0 & 32 & 32 & 32 & 64 \\
8 & 16 & 32 & 32 & 16 & 16 & 32 & 32 & 16 & 32 & 32 & 0 & 64 & 32 & 32 \\
8 & 32 & 16 & 16 & 32 & 32 & 16 & 16 & 32 & 32 & 32 & 64 & 0 & 32 & 32 \\
8 & 32 & 32 & 16 & 16 & 32 & 32 & 16 & 16 & 64 & 32 & 32 & 32 & 0 & 32 \\
8 & 32 & 16 & 32 & 16 & 32 & 16 & 32 & 16 & 32 & 64 & 32 & 32 & 32 & 0 
\end{array}
\right)
\]
\caption{
Number of Pauli terms contributing to the Hamiltonian matrix elements $H_{ij}$ and overlap matrix elements $N_{ij}$ for the $^{38}\mathrm{Ar}(0^+)$ state using the Jordan--Wigner mapping. Compared to the GC representation, the JW mapping produces substantially larger Pauli operator expansions, reflecting the reduced sparsity of the JW encoding.
}
\label{tab:gatecountsforJW}
\end{table*}

\end{appendix}

\end{document}